\documentclass[a4paper,12pt]{article}
\pdfoutput=1
\usepackage{jheppub}
\usepackage[T1]{fontenc}
\usepackage{amsfonts}
\usepackage{amsmath}
\usepackage{amssymb}
\usepackage{multirow}
\usepackage{graphicx}
\usepackage{mathtools}
\usepackage{comment}
\usepackage{graphicx}
\usepackage{subcaption}

\newcommand{\nn}{\nonumber}

\newcommand{\M}{{\cal M}}

\newcommand{\R}{\mathbb R}

\newcommand{\C}{\mathbb C}
\newcommand{\Z}{\mathbb Z}

\def\be{\begin{equation}}
\def\ee{\end{equation}}
\def\bea{\begin{eqnarray}}
\def\eea{\end{eqnarray}}

\newcommand{\Tr}{{\rm {Tr}}}

\newcommand{\f}{\frac}

\newcommand{\la}{\langle}

\numberwithin{equation}{section}

\def\la{\label}
\def\nref#1{(\ref{#1})}
\def\half{{1 \over 2 }}

\begin{document}

\title{Wormholes and the imaginary distance bound
}

\author[a]{Juan Maldacena,}

\author[b,c]{Alexander Maloney,}

\author[b,c]{and Brian McPeak}

\affiliation[a]{Institute for Advanced Study, Princeton, NJ, USA}
\affiliation[b]{Institute for Quantum \& Information Sciences, Syracuse University, Syracuse, NY, USA}
\affiliation[c]{Department of Physics, Syracuse University, Syracuse, NY, USA}

\date{\today}

\abstract{ Some of the simplest wormhole solutions involve massless scalar fields that take imaginary values. Massless fields can be interpreted as coupling constants in asymptotically flat or asymptotically AdS gravity theories. We argue that wormhole effects imply an imaginary distance bound, an upper limit for the analytic continuation of the theory to imaginary values of these couplings. In string theory examples, we find explicit effects that render the low-energy theory invalid either before or precisely at this wormhole limit. We argue that the existence of such effects enforcing the distance bound is a general feature of string theories containing wormholes. In some cases, the bounds we discuss coincide with the weak gravity conjecture, and with the Kontsevich-Segal-Witten condition on complex metrics. 
}

\maketitle

\section{Introduction}

Spacetime wormholes have played a central role in recent developments in quantum gravity \cite{Saad:2018bqo,Penington:2019kki,Almheiri:2019qdq}.  They also raise a puzzle, in that their interpretation as an average over couplings \cite{Coleman:1988cy,Giddings:1987cg,Saad:2019lba} seems to conflict with the absence of fundamental couplings in string theory. In this paper, we revisit a simple class of wormholes involving scalar fields and argue that their existence implies an absolute upper bound on how far couplings can be analytically continued in any consistent theory of quantum gravity. In certain string theory examples, we will identify non-perturbative effects 
that cause the theory to break down at or before this upper bound.
 
A simple class of Euclidean wormhole solutions can be constructed in $D$ dimensions from Einstein gravity coupled to a collection of real, massless scalar fields $\varphi_a$.  If one
analytically continues some of the scalar fields to purely imaginary values, this has the effect of effectively flipping their sign in the stress tensor and supporting a wormhole geometry \cite{Giddings:1987cg}. As we move from one end of the wormhole to the other, the scalars $\varphi_a$ shift by an imaginary amount.  Such wormholes have been extensively discussed in cases where some of the scalars are axions and can be rewritten in terms of a dual form field, which is real in the wormhole solution. We will consider the general case in which either periodic or non-periodic scalars or pseudoscalars are made complex.   

We will take the imaginary scalar fields seriously and propose an application for these wormholes.   
Consider the Dirichlet problem, where the values of the scalars are fixed at each end of the wormhole. These asymptotic values can be interpreted as coupling constants of the gravity theory.  In AdS, they are the coupling constants of the dual boundary CFT. We then consider a deformation of the theory to imaginary values of the couplings. We will interpret the wormholes as signaling a breakdown of the analytic continuation in these coupling constants. Our bound -- which we call the ``Imaginary Distance Bound'' (IDB) -- is an upper bound on how far we can analytically continue the $\langle \varphi_a \rangle$ before the theory breaks down.  

\begin{figure}[h]
    \begin{center}
    \includegraphics[scale=.35]{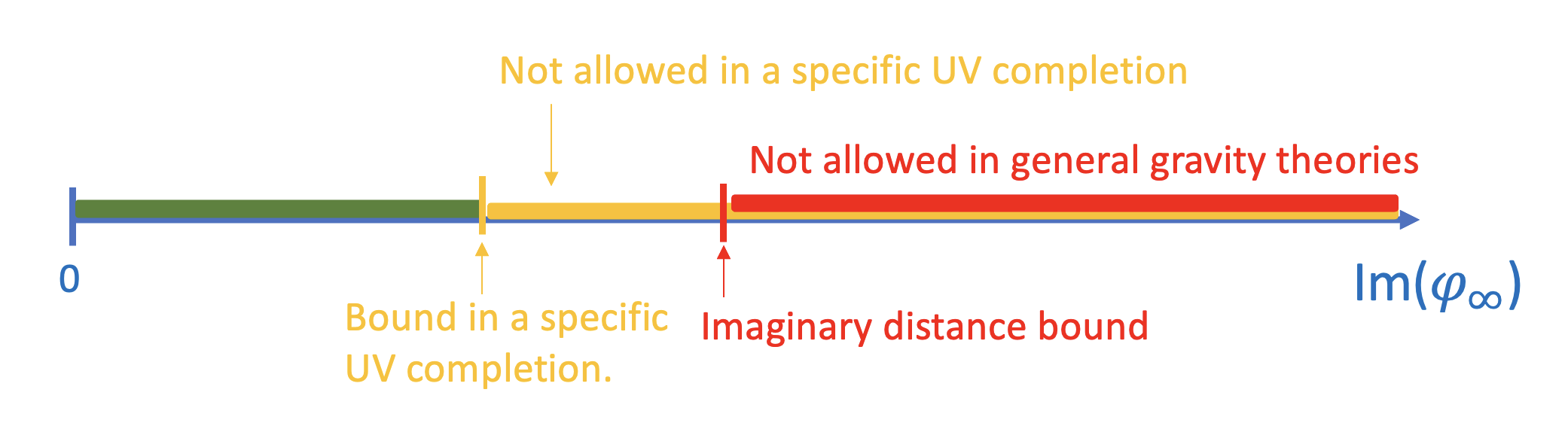}
    \end{center}
    \caption{We continue the asymptotic values $\varphi_\infty$ of scalar fields in an imaginary direction. The imaginary distance bound is the value at which an on-shell flat space wormhole appears. We propose that this places an upper bound on the analytic continuation of the coupling in any consistent theory. A specific UV complete theory (a string background) could have a lower natural analytic continuation boundary.    }
    \label{FlatAllowed}
\end{figure}

We conjecture that, in any given UV complete theory of gravity, there is always an effect intrinsic to the theory that enforces the bound.
For example, string theory has many massless scalar fields, so it supports wormhole solutions of this type. What we find is that, in the examples we have studied, there is typically some effect that renders the theory invalid at -- or in many cases before -- we get to the imaginary distance bound.\footnote{
See also \cite{Saad:2021rcu} for related ideas. }  This is depicted in figure \ref{FlatAllowed}. 

It is important to note that, although wormholes describe correlations between two different theories (the two asymptotic ends of the wormhole), we are describing a bound obeyed by a {\it single} theory.  The reason is that the exact solutions described above are good approximations to wormholes that connect two widely separated regions of the same space-time.  So their existence places limits on the analytic continuation of any individual theory.

Roughly, the physical origin of the bound is easy to understand. The idea is that the theory contains effects of the form $c_n e^{i n \varphi}$ that are bounded for real fields but could become large when $\varphi$ is imaginary. Examples of such effects are well-known in string compactifications; we will discuss several examples. In cases where these effects become important before we get to the imaginary distance bound, then the wormhole solution is rendered invalid due to these effects. This includes examples discussed by Arkani-Hamed, Orgera and Polchinski in \cite{Arkani-Hamed:2007cpn}.
In cases where the effects become important precisely at the imaginary distance bound, the wormhole leads to effects comparable in size to the corrections of the form $c_n e^{i n \varphi}$.  In these cases,  wormholes  can play a role by giving the typical value of these corrections under a suitable averaging or coarse-graining procedure.  This includes the double-cone wormhole of Saad, Shenker and Stanford \cite{Saad:2018bqo}.   We expect that such wormholes should be interpreted as genuine solutions to the UV complete theory, albeit ones that only contribute to the computation of suitably coarse-grained observables \cite{Saad:2021rcu}.  

Massless scalar fields also arise in Kaluza-Klein reductions of high-dimensional theories of gravity.  In these cases, the Imaginary Distance Bound has an important physical interpretation.  The simplest case is Kaluza-Klein reduction on a circle from $D+1$ to $D$ dimensions. The scalar field is the radius of the extra circle, and is non-compact. In this case, the wormhole in $D$ dimensions is essentially the same as the  Schwarzschild solution in $D+1$ dimensions. More precisely, it is the double cone wormhole \cite{Saad:2018bqo} associated with the Schwarzschild solution. In this particular case, the imaginary distance bound mentioned above becomes the same as the Kontsevich-Segal-Witten (KSW) \cite{Kontsevich:2021dmb,Witten:2021nzp} criterion that bounds the analytic continuation of a metric. We also show that this remains true for more general toroidal compactifications.  Our interpretation is that the Imaginary Distance Bound is a generalization of the KSW \cite{Kontsevich:2021dmb,Witten:2021nzp} bound.  Specifically, it is an extension of KSW \cite{Kontsevich:2021dmb,Witten:2021nzp} that applies to all of the scalar fields describing the moduli space of vacua, rather than just those that come from the metric. 

The Imaginary Distance Bound is also related to the weak gravity conjecture.  If we consider a Kaluza-Klein reduction starting with gravity plus a gauge field in $D+1$ dimensions, then the resulting wormholes are related to charged black holes. 
The requirement that the theory contains 
effects that dominate over the wormhole is satisfied when the $(D+1)$-dimensional theory contains matter that obeys the weak gravity conjecture \cite{Arkani-Hamed:2006emk}. In other words, the Imaginary Distance Bound is also a generalization of the weak gravity conjecture. It is closely related to the axion weak gravity conjecture in the form proposed in \cite{Montero:2015ofa}. 
This suggests that there is perhaps a more unified framework, where the KSW bound, the weak gravity conjecture, and the imaginary distance bound are all different manifestations of a single underlying principle.

Interesting wormhole solutions were discussed in \cite{Arkani-Hamed:2007cpn}. These wormholes appear to dominate over instanton effects with the same charges. Here we revisit these solutions and show that, at least for one of them, there are other effects that render the theory invalid before we get to the complex values of the moduli that appear in the wormhole solution.   Namely, we show that there are worldsheet instantons that become actionless for complex moduli values smaller than those appearing in the wormhole. This means that the wormhole solution is not a valid solution, as there are large corrections to the semiclassical theory as we go through the wormhole. 

One important caveat is that this paper assumes that the wormhole configurations contribute to the path integral. This was discussed recently for the case of axion wormholes in \cite{Witten:2026twr}. The authors of \cite{Held:2026huj} argued that the wormhole solutions we consider do not contribute to the path integral, even for imaginary values of the scalar fields (though they would contribute for the JT gravity case \cite{Held:2026bbo}). Since the precise integration contour for the gravitational path integral is not known, one should keep in mind that the definition of \cite{Held:2026huj} leads to a different picture. The interpretation of \cite{Held:2026huj} would also suggest that the Schwarzschild version of the Saad, Shenker Stanford (SSS) double cone \cite{Saad:2018bqo} does not contribute.\footnote{The perturbative stability of wormholes has also recently been discussed, see \textit{e.g.} \cite{Hertog:2018kbz, Loges:2022nuw, Aguilar-Gutierrez:2023ril,Hertog:2024nys, Ivo:2026ijv, Loveridge:2025dls}, where the consensus seems to be that that wormholes are stable for the Neumann problem (\textit{i.e.} fixing the form-fields rather than the scalars at the boundary) and unstable for the Dirichlet problem.}

This paper is organized as follows. In section \ref{sec:solutions}, we review scalar field wormhole solutions in flat space and AdS, viewing them as solutions to a Dirichlet problem where we fix the values of the scalar fields at infinity.  We show that the action of these solutions vanishes when the scalar traverses too large an imaginary distance.
In section \ref{sec:bound}, we discuss the bound on imaginary distances. First we discuss some examples of analytic continuations of the couplings in quantum field theories, and note that there is a natural boundary for these analytic continuation. We then use the wormholes to motivate the imaginary distance bound, first in flat space and then in AdS space. 
In section \ref{sec:dimensional}, we consider examples of theories arising from Kaluza-Klein reduction. We discuss the connection to the SSS wormhole \cite{Saad:2018bqo},  the KSW criterion and the weak gravity conjecture. 
In section \ref{sec:UV}, we discuss some examples of wormholes in UV complete theories defined via string theory.  In several examples, we show that the bound is obeyed by displaying effects that act as a natural boundary for analytic continuation prior to the Imaginary Distance Bound.

  {\bf Note:} While this paper was in preparation, we learned that the authors of \cite{OtherPaper} were thinking in the same imaginary direction. The submission of this paper is coordinated with that of \cite{OtherPaper}; at certain points we will refer the reader \cite{OtherPaper} for issues that are better handled by their approach.

\section{Wormhole solutions}
\label{sec:solutions}

\begin{figure}[h]
    \begin{center}
    \includegraphics[scale=.4]{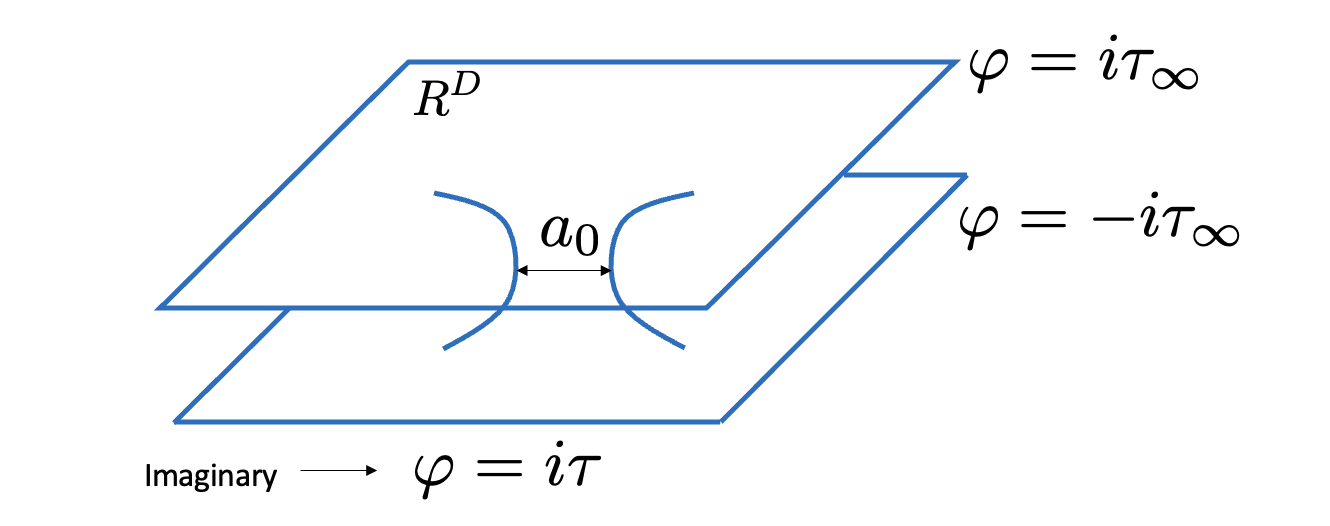}
    \end{center}
    \caption{Sketch of the wormhole solution  }
    \label{WormholeFig}
\end{figure}
 
We begin in this section by reviewing the wormhole solutions of Euclidean $D$-dimensional gravity coupled to a massless scalar field:
\def\grcou{16\pi G_N} 
\be \la{SingFi}
I =   { 1 \over \grcou } \int d^Dx \sqrt{g} \left[-( R - 2 \Lambda)  + \half (\nabla \varphi )^2 \right] ~.
\ee 
The scalar field is dimensionless, and is canonically normalized in Planck units.  Note 
that we keep the standard positive sign for the kinetic term. The precise normalization of the scalar actions chosen here will be important. 
More generally, we can consider several scalar fields 
\be \la{ManFi}
I =   { 1 \over \grcou } \int d^Dx \sqrt{g} \left[ -(R - 2 \Lambda)  + \half  h_{ab}(\varphi) \nabla \varphi^a  \nabla \varphi^b \right] ~,
\ee 
where $h_{ab}$ is the metric in the scalar field target space, sometimes called moduli space. In the dual CFT, the asymptotic values of the $\varphi^a$ are couplings, and $h_{ab}$ is the Zamolodchikov metric on the space of couplings. This moduli space need not have any isometries or shift symmetries. We now discuss wormhole solutions in flat and AdS space.\footnote{There are also wormhole solutions in de Sitter space which we will not discuss here, because the Dirichlet problem -- the main interest of this work -- is harder to set up in that case.}

\subsection{Wormhole solutions in flat space }

We start with $\Lambda =0$ and write a  spherically symmetric ansatz of the form
\be \la{MeWH}
ds^2 = { d a^2 \over  F(a)   } + a^2 d\Omega_{D-1}^2 ~,~~~~~~~~~ \varphi = i \tau(a) \, .
\ee 
A solution of \nref{SingFi} (with $\Lambda =0$) is \cite{Giddings:1987cg} 
\be
\la{FaWH}
F(a) = 1 - \left( { a_0 \over a } \right)^{ 2 (D-2)}   ,~~~~~\tau = \sqrt{ 2 (D-1) \over (D-2) } \tilde \tau  ~,~~~~~~ \cos \tilde \tau = \left( { a_0 \over a } \right)^{   D-2}  \, .
\ee 
We can think of $\tau$ as defining an imaginary displacement of the scalar field, which moves the scalar an imaginary distance in moduli space\footnote{Note that the Euclidean path integral is over {\it real} values of the scalar fields $\varphi$, regardless of whether these fields are axions or not. See a discussion in \cite{Witten:2026twr}.  Here we are focusing on an imaginary solution because we are choosing imaginary boundary conditions for the field.}. Alternatively, we can think of $\tau$ as a timelike coordinate in a Lorentzian continuation of the target space. 
The square root branch cut in the function $\tau(a)$ at $a\sim a_0$  implies that our solution has two sheets, corresponding to the two sides of the wormhole; see Figure \ref{WormholeFig}. 
$a_0$ is the size of the ``waist" of the wormhole, \textit{i.e.} the minimum size of the sphere. The field displacement as one moves from one side of the wormhole to the other is independent of $a_0$, and is
\be \la{Diff}
-i \Delta \varphi_\infty = \Delta \tau_\infty  = \pi \sqrt{ 2 (D-1) \over (D-2) }
\ee 
The action \nref{SingFi} has a shift symmetry for $\varphi$, so we can add any constant to $\varphi$. We will normally fix these constants so that $\varphi$ is real at $a=a_0$.

We can consider the Dirichlet problem where we fix the value of the field at the two asymptotic regions. For generic asymptotic values the solution does not exist. However, for the specific values obeying 
\nref{Diff} the solution does exist. Moreover, there is a one parameter family of solutions characterized by $a_0$. We can easily evaluate the action of these solutions. The trace of Einstein's equations gives 
$R - \half (\nabla \varphi)^2 =0$, which implies that the on shell action is zero,
\be \la{ZeroFlat}
I =0 
\ee 
There is no contribution from the  Gibbons-Hawking term because the metric approaches that of flat space sufficiently quickly. 

These features follow from a scaling similarity of the action \nref{SingFi} when $\Lambda=0$. Namely, rescaling the metric only rescales the action by an overall constant. As a result, the equations of motion are invariant; so if there is a solution, it can have any size. This means that the action can only be zero -- if it had not been, it would depend on $a_0$  as $a_0^{D-2}$, and therefore would not be an extremum with respect to variations of $a_0$. This can happen only for special boundary conditions obeying \nref{Diff}. 

We will mostly discuss this Dirichlet problem. It is also common in the literature to consider the problem where the wormhole charge is fixed. This is the charge under the global symmetry of $\varphi $ translations, 
\be \la{Charge}
q = -i  \int_{S^{D-1} } \partial_n \varphi 
= { 1 \over 16 \pi G_N} \omega_{D-1} \sqrt{ 2 (D-1)(D-2) } a_0^{D-2}
\ee 
For this fixed-charge problem, the action is non-zero and equal to the Legendre transform of the action for the Dirichlet problem\footnote{In the full partition function, this is a Fourier transform, see \cite{OtherPaper} for more discussion. } 
\be 
I_q = -i   \Delta \varphi_\infty q 
\ee
where $ -i \Delta \varphi_\infty$ is given in \nref{Diff} and is real.  We have used the fact that the action for the Dirichlet problem is zero. It is common to discuss wormholes in the context of compact fields, including axions. We will consider both compact and non-compact fields. 

A related computation is the following. Suppose that we start from a wormhole configuration of a certain size $a_0$. This is a solution when the boundary conditions satisfy \nref{Diff}. Let us now change the boundary conditions slightly and ask how the action changes. This is given again by the conjugate momentum as 
\be \la{ActFlatOff}
\delta I = -i \delta \varphi_{\infty} q \propto - a_0^{D-2} \delta \tau_{\infty }
\ee 
This means that the variable $a_0$ gets a ``potential'' which would drive it to $a_0\to 0$ if $\delta \tau_{\infty} $ is negative, and to infinity if $\delta \tau_{\infty}$ positive. This shows that the special value \nref{Diff} is a transition point between the two behaviors. 

\subsection{Wormhole solutions in Euclidean AdS$_D$ (or $H_D$)   }

We now set $2 \Lambda = -   (D-1)(D-2)/L^2$. 
With the same ansatz \nref{MeWH} we find the solution 
\bea 
F(a) &=& 1 + { a^2 \over L^2} - \left( 1 + { a_0^2 \over L^2 } \right) \left( {a_0 \over a } \right)^{2(D-2) }
~,~~~
\cr 
 \tau(a) &=& \sqrt{ 2 (D-1)(D-2) \left(1 + {a_0^2 \over L^2 } \right)} \int_{a_0}^{a} { d \tilde a \over \sqrt{F(\tilde a ) }} {a_0^{D-2} \over {\tilde a}^{D-1}  }
\la{AdSSol} \eea 
For small values of $a_0$ we recover the flat space wormhole solution and the asymptotic values of $\tau$ obey \nref{Diff}. 

\begin{figure}
 \begin{center}
    \includegraphics[scale=.6]{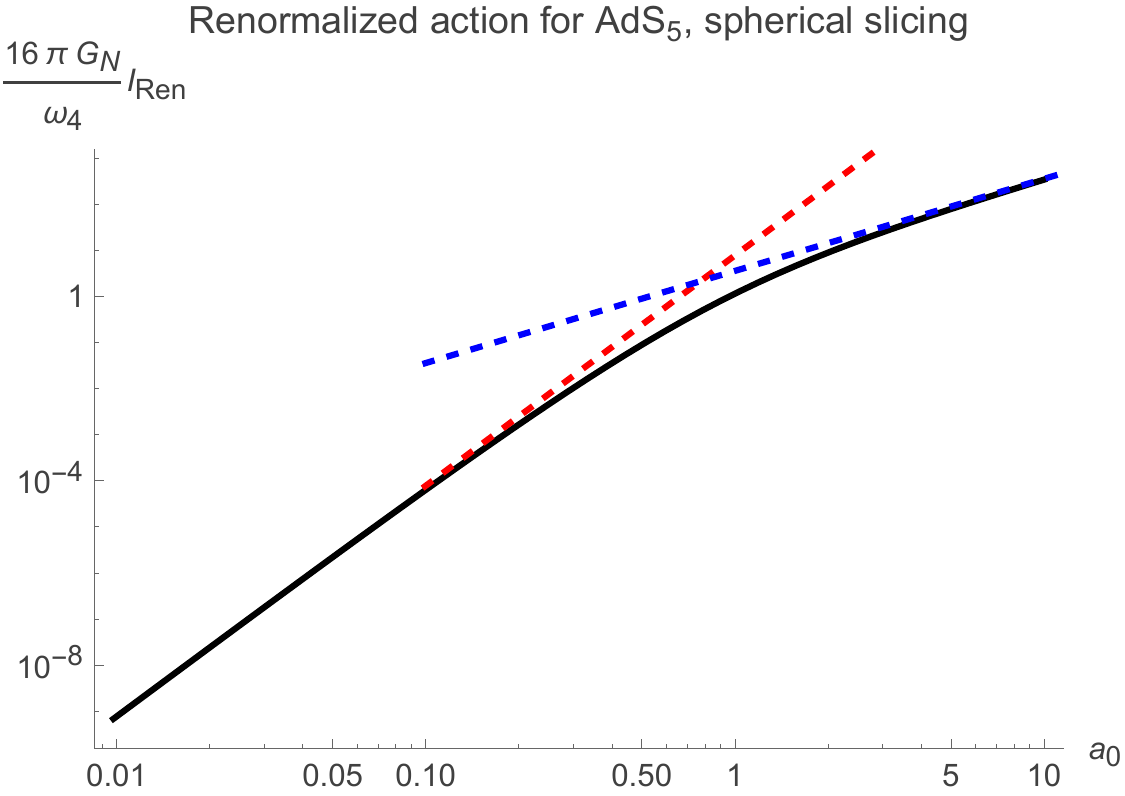}
    \end{center}
    \caption{Log-Log plot of the action for the AdS$_5$ wormholes showing also the small $a_0$ (dotted red line) \nref{SmallAs} and large $a_0$ (dotted blue line) asymptotics \nref{BigAs}. We set $L=1$. }
    \label{fig:Iren}
\end{figure}

We can evaluate the action as follows. First, note that the trace of Einstein's equations imply 
$R - \half (\nabla \varphi)^2 =   2 D \Lambda/(D-2)$, which implies that the action is proportional to the volume, $I \propto \int \sqrt{g}$
\be \la{AdSAct}
I_{\rm ren} = { 2 (D-1) \over \grcou } 2 \omega_{D-1} \left[ \int_{a_0}^\infty da  { a^{D-1} \over \sqrt{F(a) } } - \int_0^\infty da  {  a^{D-1} \over \sqrt{ 1 + a^2/L^2 } }  \right] \, ,
\ee 
 where we subtracted the volume of AdS$_D$ and $\omega_{D-1}$ is the area of the sphere $S^{D-1}$. The Gibbons-Hawking term is the same as that of empty AdS$_D$  because the metric approaches the vacuum rapidly. One can check numerically that the action is positive, see figure \ref{fig:Iren}. Moreover, for small and large $a_0$ we have  
\bea \la{SmallAs}
     I_\text{ren} &\simeq &\frac{\omega_{D-1}}{16 \pi G_N} \frac{4\sqrt{\pi}(D-1)}{D(D-2)}
\frac{\Gamma\!\left(\frac{D-4}{2D-4}\right)}
{\Gamma\!\left(\frac{D-3}{D-2}\right)}
\,  { a_0^{D} \over L^2} ~,~~~~~~~~~~~~~~~~~ { a_0 \over L } \ll 1 
 \\
\la{BigAs}
     I_\text{ren} &\simeq & \frac{\omega_{D-1}}{16 \pi G_N} \frac{4\sqrt{\pi}\,(D-2)}{(D-3)}
\frac{\Gamma\!\left(\frac12+\frac{1}{D-1}\right)}
{\Gamma\!\left(\frac{1}{D-1}\right)}
\,a_0^{D-3} L \,  ~,~~~~~~~~ { a_0 \over L } \gg 1 
\eea

For large $a_0$ the solution goes over  to the solution for flat spatial slicing, which takes the form\footnote{ Note that in \nref{PoinAct} we can replace $d{\vec x}^2 $ by any Ricci flat manifold. For example, we can have a $D=5$ solution where the boundaries are a K3. We can further discuss solutions where the moduli of the K3 vary, since they will behave as additional massless fields.} 
\bea \la{PoinAct}
ds^2 &=& L^2 \left[   { d \hat a^2 \over  F(\hat a)   } + a_0^2 \hat a^2 d {\vec x}^2 \right] ~,~~~~~F(\hat a ) = \hat a^2 \left( 1 - { 1 \over {\hat a }^{ 2 (D-1) } } \right) 
\cr 
\tau &=& \check \tau \sqrt{ 2 (D-2) \over (D-1) } ~,~~~~~~ \cos \check \tau = { 1 \over {\hat a }^{D-1} }
 ~,~~~~~~~I_{\rm ren} =0\eea 
The action for this solution is zero. 
This solution spontaneously breaks the scaling symmetry. Actually, the problem has two scaling symmetries, one for each side. Both are broken -- $a_0$ is the Goldstone boson for the overall scaling, and there is a second that arises from relative rescalings of the two boundaries.

One might be confused by the statement that the action \nref{AdSAct} grows with $a_0$  while \nref{PoinAct} is zero. The point is that the zero action in \nref{PoinAct} should be interpreted as the vanishing of the coefficient of the naively leading power in $a_0$, going like $a_0^{D-1}$ in \nref{BigAs}. 

In  \nref{PoinAct},  the displacement of the scalar field along the imaginary direction is 
\be 
\Delta \tau_{\infty}' = \pi  \sqrt{ 2 (D-2) \over (D-1) }
\ee 
which is smaller than \nref{Diff}. For general values of $a_0$ the AdS$_D$ solution \nref{AdSSol} has an imaginary difference which is between the two imaginary asymptotic  values 
\be \la{RangeDel}
 \pi  \sqrt{ 2 (D-2) \over (D-1) }= \Delta \tau_{\infty}' < \Delta \tau_\infty(a_0)  <  \Delta \tau_{\infty} =\pi  \sqrt{ 2 (D-1) \over (D-2) }
 \ee 
The two extreme values correspond to $a_0=0$ and $a_0=\infty$. 

For small $a_0$, a simple way to understand why there is a solution for a range of values is the following. In flat space, we have seen that a boundary value of the scalar field slightly away from the critical one leads to the off-shell action \nref{ActFlatOff}. If we now add a negative cosmological constant,  we get an extra term that comes from the volume of the wormhole. We then get the off-shell action 
\be 
\la{ActAdSOff}
I = - a_0^{D-2} \delta \tau_{\infty} - c |\Lambda | a_0^{D} ~,~~~~~~~c>0 ~,~~~~~~~{a_0 \over L } \ll 1
\ee 
For $\delta \tau_{\infty} <0$, there is a solution for a non-zero $a_0$ but we see that the solution is a local maximum as a function of $a_0$, where we recover \nref{SmallAs}.

\subsection{Solutions with many scalar fields}

When we have a more complicated moduli space \nref{ManFi}, which is reflected in the sigma model metric $h_{ab}(\varphi)$, the wormhole solutions have exactly the same geometry 
\nref{MeWH} and \nref{FaWH}.  The only difference is that the equations of motion for the scalar fields imply that the moduli $\varphi^a$ will trace out a geodesic through moduli space.  The $\tau(a)$ appearing in the solution is the proper imaginary distance along this geodesic \cite{Arkani-Hamed:2007cpn}. This is true both for the flat space and the AdS$_D $ wormholes. 
We will be interested in geodesics with a purely imaginary proper distance. 
The solution then describes a portion of a geodesic whose proper distance is again \nref{Diff} in the flat space case, and falls in the range \nref{RangeDel} for the AdS$_D$ case. The geometry is exactly the same as that of the single field case. 

The main novelty in the multi-field situation is that we now have motion on a curved moduli space. In principle, this curved moduli space need not have any symmetries and we could consider displacement along imaginary distances along an arbitrary complexified geodesic. The simplest two-field case we will consider is one where the moduli space has the geometry of hyperbolic space\footnote{This system was studied in a series of papers \cite{Bergshoeff:2004fq, Bergshoeff:2004pg, Bergshoeff:2005zf}, where, in addition to wormholes, the authors discussed ``extremal'' and ``superextremal'' instanton solutions, which are classified on the basis of their $SL(2, \mathbb R)$ charges.}
\be 
h_{ab} d\varphi^a d\varphi^b = \kappa^2 \left( { d Z^2 +  d C^2 \over Z^2 } \right) 
\ee 
where $\kappa$ sets the radius. The imaginary displacement could be along any direction. For example, we can set $C=0 $ and move along an imaginary $ \log Z $ direction. 

\begin{figure}
 \begin{center}
    \includegraphics[scale=.4]{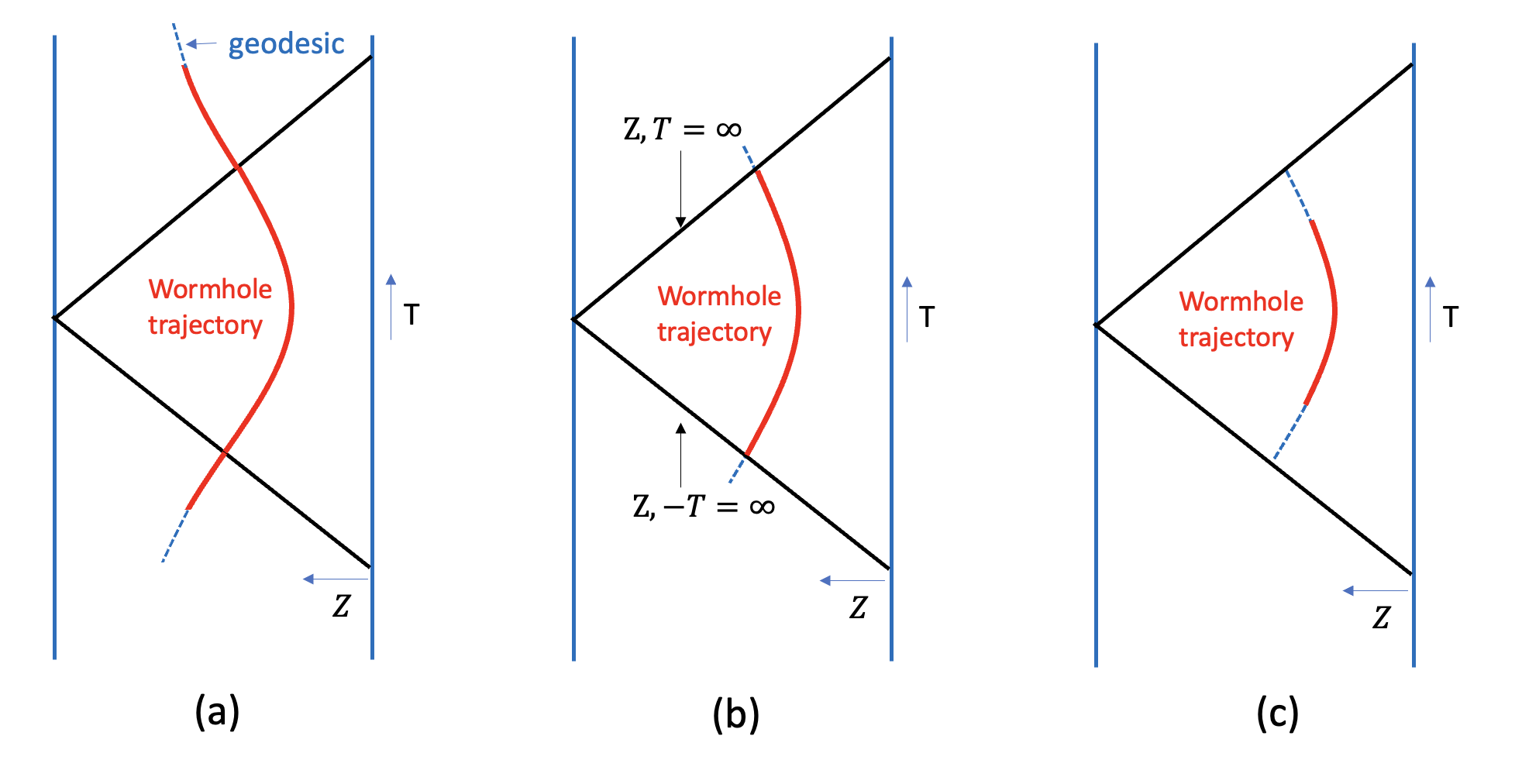}
    \end{center}
    \caption{In wormhole solutions the scalar fields lie along geodesics. In the simple case that the moduli space has been complexified to AdS$_2$, depending on the overall scale of the moduli space, the portion of the geodesic covered by the wormhole can cover different portions of the geodesic. All are denoted in red and have the same proper length. The three plots differ by the overall radius of the AdS$_2$ (or original $H_2$) space. This radius increases from left to right, from (a) to (c).  }
    \label{Geodesics}
\end{figure}

 Alternatively, we could take our solution to have real fields with a wrong sign kinetic term. Then the moduli space would appear Lorentzian (or have mixed signature more generally), and the wormholes would correspond to timelike geodesics on moduli space. In the example above this would mean that we set $C = i T$ and to look for a timelike geodesic in the metric 
\be \la{AdS2Po}
ds^2 = \kappa^2   \left( { d Z^2 -  d T^2 \over Z^2 } \right)
\ee 
This has the geometry of AdS$_2$ and the geodesics are the oscillatory curves shown in figure \ref{Geodesics}.

For example, a simple family of geodesics, parametrized in terms of proper time, $\tau$, is
\be  \la{TrajGeo}
  {  Z }  = e^\gamma { 1 \over \cos(\tau/\kappa) } ~,~~~~~~~~~~ - i C = T = e^{ \gamma } \tan (\tau/\kappa) 
  \ee 
 Here $\gamma$ is an integration constant related to the choice of the geodesic. We could also add another integration constant by replacing $T \to T + T_0 $ for some constant $T_0$.  We could introduce another parameter by simply adding a constant to $\tau$. 
 For this discussion, the important property of geodesics in AdS$_2$ is that the proper time that they spend in the Poincar\'e patch -- the region with $Z>0$ -- is
 \be 
  \Delta \tau_{\rm Poinc} = \pi \kappa 
  \ee 
  and is independent of the choice of geodesic. 

This means that if we want a geodesic which stays within the Poincar\'e patch $Z>0 $ then we would require that $\Delta \tau_{\text{Poinc}} $ is larger than $\Delta \tau_\infty $ for the wormhole in question. 
So the value of $\kappa$ becomes important. We will discuss various examples of moduli spaces below. One simple example is the axi-dilaton of the ten-dimensional type IIB supergravity, which has $\kappa=1$. 

Note that if we are analytically continuing in the coupling constants, there is nothing special about the $Z>0$ region. We can certainly continue beyond this region, and consider any complex extension of the moduli space. In particular, coordinate singularities such as the one at $Z=\infty$ in \nref{AdS2Po} can always be removed by the field redefinition which corresponds to going to global coordinates on moduli space. In other words, the ``horizon'' at $Z=\infty $ does not mean anything special from the point of view of the low-energy gravitational theory; of course, this might change in the full UV complete theory. We discuss this more in a particular example in section \ref{sec:UV}.

\section{A bound on imaginary distances}
\label{sec:bound}

The asymptotic values of massless fields $\varphi^a$ can be viewed as coupling constants. For example, the expectation value of the dilaton is related to the string coupling constant.  In AdS, these asymptotic values are the couplings of the boundary conformal field theory, and $h_{ab}$ is the Zamolodchikov metric on corresponding moduli space of conformal field theories.
It is natural to ask whether we could deform such a theory, either in flat space or AdS, to complex values of the couplings.  Of course, this will ruin the unitarity of the Lorentzian version of the theory, since we would have a Euclidean theory which is not reflection positive. Such Euclidean theories can be of interest as statistical mechanics problems. 

 The wormholes signal a breakdown of the theory when we have two asymptotic regions separated by imaginary distances greater than $\Delta \tau_\infty$ or $\Delta \tau'_\infty$ in moduli space. This motivates our main conjecture, which is that the theory with a single boundary stops making sense when we go beyond half this distance from the real line. Here, by ``making sense'' we simply mean that the low-energy effective field theory described by Einstein gravity with massless (or approximately massless) fields remains a valid approximation. In this section we formulate the exact conjecture and give some heuristic arguments; the rest of the paper will be devoted to exploring other evidence.

\subsection{A field theory example with imaginary couplings}
\label{sec:ftex}

Before considering wormholes, let us investigate some ways in which a theory could stop making sense as we analytically continue in some couplings. 
As an example, consider the case of four-dimensional ${\cal N} =4 $ super Yang-Mills theory. This theory has two coupling constants, the gauge coupling and the $\theta$ angle. The theory certainly makes sense when both couplings are real. 
We can imagine taking the $\theta $ angle to have some imaginary value, so that $i\theta$ is real. The euclidean path integral continues to be well defined, at least when
\be  \la{BoundTh}
 |i \theta| < { 4 \pi^2 \over g^2_{YM} } 
 \ee 
 Each additional instanton would contribute with an extra factor of 
 \be \la{QDef}
  q = \exp\left( -{ 4 \pi^2 \over g^2_{YM} } + i \theta \right) 
  \ee 
  When $i\theta$ approaches the upper bound we find that the inequality \nref{BoundTh} is reversed and $q>1$, and also the classical action becomes unbounded below. Therefore we expect that the theory would ``not make sense''. 

  More explicitly,    Vafa and Witten computed the partition function of a twisted version of ${\cal N}=4$ super Yang-Mills \cite{Vafa:1994tf}. For the case of K3 and gauge group SU(2),  they found\footnote{They also found results for SU(N),   in the sense that they diverge as $q\to 1$.}
  \be \la{VWPart}
  Z(\text{K3}) = {1\over 8 } G(q^2) + { 1 \over 4 } G( q^{1/2} ) + { 1 \over 4 } G( - q^{1/2} ) ~,~~~~~~~G(q) \equiv  { 1 \over \eta(q)^{26 } } ~,~~~~~~~~ \eta \equiv q^{  { 1 \over 24} } \prod_{n=1}^\infty (1 - q^n) 
  \ee 
 with $q$ in \nref{QDef}. $Z(\text{K3})$ blows up when $q\to 1$. In fact, the singularity is so bad that it is not possible to continue the function beyond $|q|=1$. Indeed, classic results in complex analysis going back to Polya state that generic analytic functions with radius of convergence 1 cannot be analytically continued outside of the unit disc. The familiar examples of functions that admit analytic continuations outside of their radius of convergence all require some special structure, typically associated with integrability.

 We have discussed continuing in $\theta$. But we could also discuss continuing to complex values of the Yang-Mills coupling $g^2_{YM}$. For example, we could change its phase, $g^2_{YM} \to g^2_{YM} e^{ i \varphi} $. From \nref{VWPart}, we see that the expression becomes singular when $e^{ i \varphi } = i $ and $|q|=1$ (now for real $\theta$).  More generally, there will always be a problem when $\text{Im} (-{ 4 \pi^2 \over g^2_{YM} } + i \theta) < 0$.  In the bulk string theory, this corresponds to $D(-1)$ instantons becoming actionless, as we discuss in section \ref{sec:UV}.

\subsection{Gravity at imaginary values of the couplings}

Let us now consider gravitational theories, where the couplings are the asymptotic values of massless scalars. Even when the theory makes sense for real values, it might be that it does not make sense as a Euclidean problem at imaginary values of the couplings. Again, we emphasize that by ``not making sense'' what we mean is that the semiclassical gravitational theory that we started with would no longer be a good approximation to the theory at these imaginary couplings.

A simple reason why the theory might not make sense is the following. Imagine that there are some terms in the action that are small for all real values of the coupling, such as a term 
\be \la{Corr}
c_n e^{ \pm i n \varphi }
\ee 
where $n$ could be a discrete or continuous label. If $c_n$ is sufficiently small, then we can neglect such terms for real $\varphi$. However, if $\varphi = i \tau $ for a real $\tau$, then such terms can become large, and when  $c_n e^{ n \tau } \sim 1 $ they cannot be neglected, and the low-energy theory \nref{SingFi} is strongly modified. In order to figure out how precisely it is modified, we would need to compute such terms. In a well defined theory of quantum gravity, such terms could have various origins, including perturbative string corrections, $D$-instantons, etc. These effects can originate independently of any wormholes. These corrections could appear in a potential term, or in some higher-derivative term, as will be the case in supersymmetric theories where the wormhole has some zero modes that need to be absorbed. But in any case, the idea is that the theory fails to make sense because there are some corrections that are becoming large. 

Let us now consider the wormholes we discussed above. For the flat space case, these wormhole solutions only connect vacua with different complex values of the couplings, at least for on-shell solutions. 
The presence of the on-shell wormholes, which exist for any value of $a_0$, seems to signal the onset of an instability
when we have a pair of flat space theories with values of the coupling that are separated by a special imaginary distance in moduli space \nref{Diff}. One interpretation would be to say that some average over couplings diverges when the imaginary distance between the two is \nref{Diff}. 

We would like to make a stronger conjecture.  To motivate it, let us start with the theory with real couplings, or real asymptotic vacuum expectation values for the fields\footnote{ Note that this conjecture requires that there is a distinguished real line to start from. In particular, it would not make sense if the scalar has an exact shift symmetry. It is expected that such a global symmetry would be broken in quantum gravity, but it is indeed an additional assumption.}. We then start analytically continuing the theory to complex vacuum expectation values along an imaginary direction. We propose that the theory cannot  ``make sense'' for  imaginary distance beyond the range  
 \be \la{IDB}
\boxed{ ~~ \text{Flat space IDB}: ~~~~~~~~~~\tau \leq \tau_{\rm{IDB}} ~,~~~~~~~~~~~\tau_{\rm IDB} \equiv { \pi \over 2 } \sqrt{ 2 (D-1) \over (D-2) } ~~ }
\ee  
This is what we will call the flat space ``Imaginary Distance Bound,'' (IDB). Notice that $\tau_{\rm{IDB}}$ is {\it half} of the imaginary distance traveled by the wormhole solutions (\textit{i.e.} $\tau_{\rm{IDB}} = \Delta \tau_{\infty} / 2$) because we are considering a solution where the scalar is real at the midpoint of the wormhole.   Note that this displacement is unitless in our conventions; see appendix \ref{app:axions} for the units when the scalars are axions.
 
If we imagine starting from some point $p$ in moduli space and moving a distance $i \tau $ in the imaginary direction, then we can compute 
\be \la{PairZ}
Z(p + i \tau ) Z(p - i \tau ) 
\ee 
This  receives contributions from on-shell wormholes when $\tau $ reaches the critical value \nref{IDB}. The fact that these contributions have an arbitrary value of $a_0$ suggests that we are reaching a point where they will start diverging. Namely, that they would diverge for $\tau > \tau_{IDB}$\footnote{This point is made more clear in \cite{OtherPaper}, which discusses the sum over fixed charged wormholes.  }. 
At first sight,  this divergence would be associated with the pair of theories in \nref{PairZ}.  
The main content of the IDB, then, is to associate the divergence with each individual theory. Namely, the idea is that there are corrections of the form \nref{Corr} that are becoming large. The only thing we can conclude from this discussion is that the whole sum (or integral) over $n$ in \nref{Corr} is becoming large. However, if we interpret the fixed charge solution in terms of average over couplings, an average over $c_n$, then we would also conclude that each term $c_n e^{ n \tau} $ is becoming of order one, and cannot be neglected. 

Another point of view is that if we have a range of imaginary distances that make sense, then we can consider any pair of theories and compute
\be \la{PairZ12}
Z(p + i \tau_1) Z( p - i \tau_2) 
\ee 
If the individual theories make sense, then the pair should make sense as well. This will be the case if both $\tau_1$ are less than \nref{IDB}. However, if one of them, say $\tau_1 > \tau_{IDB}$, then we can find a $\tau_2$, with $\tau_2 < \tau_{IDB}$ such that \nref{PairZ12} becomes large.  In other words, if both $\tau_1 , \tau_2 < \tau_{IDB}$, then \nref{PairZ12} is not very large.

 A third perspective on why there might be a single-sided bound would be to consider the wormhole to connect very distant regions in the same spacetime. If we, for instance, set up a very shallow imaginary scalar field gradient over very long distances, then there would be zero-action wormholes connecting these very distant locations. 

The bound \nref{IDB} derives from the classical wormhole solutions. One might wonder about the impact of loop corrections or other general higher-derivative corrections to the action. 
In this flat space case, the bound comes from large wormholes, with $a_0 \to \infty$, so for those solutions we can neglect such corrections. In other words, using the off-shell action for small positive deviations away from the bound given in \nref{ActFlatOff}, and including a one-loop correction we get path integral including $\int d a_0 a^p_0 \exp\left(  \delta \tau_{\infty}   a_0^{D-2} \right) $, for some power $p$. We see that the exponential term dominates for large $a_0$, so the effect of any loop correction is negligible.

\begin{figure}[h]
    \begin{center}
    \includegraphics[scale=.35]{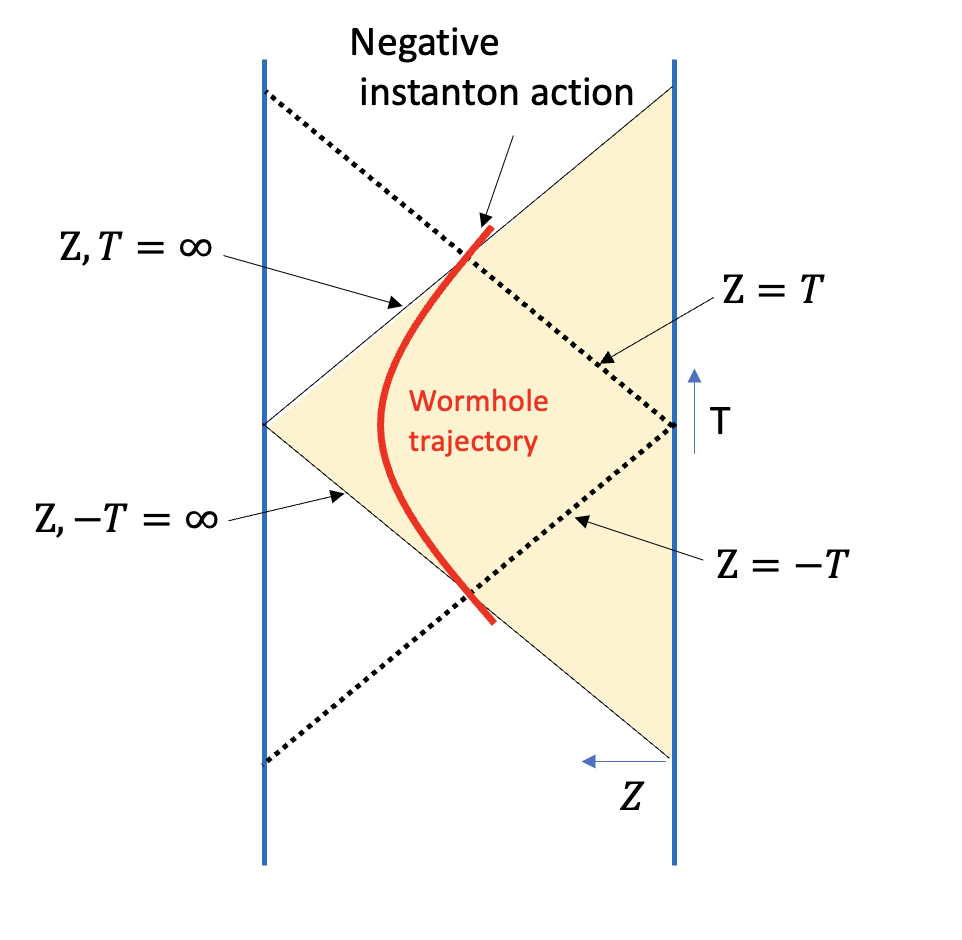}
    \end{center}
    \caption{In yellow we see the portion of a complexified slice of the moduli space parametrized by positive $Z$ and real $T$. The trajectory of a wormhole is plotted in red. This wormhole trajectory crosses the lines  $Z = \pm T $ where the instanton action \nref{InstAct} switches sign and becomes negative. This means that the original semiclassical theory becomes invalid as we cross these lines. }
    \label{AdS2Instanton}
\end{figure}

In some theories, the problem with the low-energy approximation appears {\it before } we get to \nref{IDB}. For example, if we consider type IIB in ten dimensions, then the axi-dilaton moduli space has the metric 
\be 
\label{IIBmod}
h_{ab} d\varphi^a d\varphi^b = d\phi^2 + e^{  2\phi } dC_0^2 
\ee 
  which is a hyperbolic space of radius one. The timelike proper distance to the Poincar\'e horizon is $\pi/2$, which is less than \nref{IDB} for all $D$. When we cross the Poincaré horizon, the sign of the instanton factor changes for time-symmetric trajectories (see figure \ref{AdS2Instanton}). In other words 
  \be \la{InstAct}
  I_{\rm instanton}  = e^{ - \phi } - i C_0 =Z - T  
   \ee  
   changes sign.

 \begin{equation}
\fbox{
\begin{minipage}{0.8\linewidth}
\it In a well-defined, UV complete, quantum gravity theory, there should be some effect that renders the low-energy approximation invalid either at or before $\tau_{\text{\textup{IDB}}}$, as we set the couplings to imaginary values.
\end{minipage}
}
\label{NDWC}
\end{equation}
This can be viewed as an extension of the weak gravity conjecture, 
as we discuss in section \ref{sec:dimensional}.  It can also be viewed as related to the half wormhole 
solutions of
\cite{Saad:2021rcu}, since those half wormholes would be examples of effects that are comparable to the wormhole contributions. 

We should emphasize that this conjecture applies to any moduli space; we do not require a periodic scalar (axion), nor do we require a shift symmetry. Whatever the moduli space, we complexify it, and we move an imaginary distance from any point on the real slice. Then \nref{IDB} bounds the amount we can displace the moduli. 

We also emphasize that a theory with imaginary couplings is not a physical unitary theory.  We are only bounding certain unphysical deformations of the theory. Still, it is interesting to note that we could analytically continue one of the non-compact coordinates of the wormhole to get a Lorentzian solution. This would lead to a configuration that looks like a bubble of nothing, representing the decay of two copies of spacetime with opposite imaginary expectation values for the scalar. This is not a decay process of a physical theory -- it is a decay process involving the {\it unphysical} theories with imaginary couplings. Furthermore, since the action vanishes, this decay happens very rapidly. This is reminiscent of the Witten bubble of nothing that describes the decay of circle compactification of flat space (with antiperiodic boundary conditions for the fermions) \cite{Witten:1981gj}.  But there are two important differences. First, the Witten bubble of nothing describes the decay of a physical compactification. Second,  its action is positive, indicating that the decay rate is exponentially small.

\subsection{ An AdS Imaginary distance bound }

In Anti-de Sitter space with spherical slices, if spacetime has a large radius, then we also locally expect to have the bound
\nref{IDB} for the local low-energy gravity approximation. However, we have seen that there are also wormhole solutions with asymptotic boundary values in the range \nref{RangeDel}. One important difference is that such wormholes typically have positive actions, so their contributions are small.  So if we view them as arising from random couplings, then the values of these couplings are typically small. 

In AdS, there are 
wormholes with flat slicing as well as spherical slicing in the bulk. For flat slicing, we have zero-action wormholes with arbitrary size. From equation \nref{RangeDel}, we see that these wormholes appear at a lower value of $\tau$ than the flat space wormholes. So we have the following bound:
 \be \la{IDBAdS}
\boxed{ ~~ \text{AdS IDB}: ~~~~~~~~~~\tau \leq \tau'_{\rm{IDB}} ~,~~~~~~~~~~~\tau'_{\rm IDB} \equiv { \pi \over 2 } \sqrt{ 2 (D-2) \over (D-1) } ~~ }
\ee  
The spherical wormholes are a little trickier, as the action vanishes at $\tau_\text{IDB}$ \nref{IDB} rather than the earlier point $\tau'_\text{IDB}$ given by \nref{IDBAdS}. So it is natural to wonder whether this bound \nref{IDBAdS} should also extend to the case that we have a spherical boundary. 
The authors in \cite{OtherPaper} suggested this based on the convergence of the sum over fixed charged wormholes, which is an advantage of their approach. Let us briefly explain that point of view. We first consider the wormholes for fixed charge $q$ and compute the fixed charge action $I(q)$. We then consider the sum or integral over the charges of the form 
\be \la{SumQ}
\sum_ q Z_1(q) \exp\left( -I(q) + q \Delta \tau \right) 
\ee 
where $Z_1(q)$ is a one loop factor, which is expected to be a polynomial in $q$.  
Moreover, we know that the on shell solution obeys $\partial_q I(q) = \Delta \tau(q)$ where $\Delta \tau(q)$ is the total imaginary displacement we see in the solution \nref{AdSSol}. In particular, we find that the large $q$ behavior of the fixed charge action is
$I(q) \sim  q \Delta \tau'_{\infty }$, with $\Delta \tau'_{\infty }= 2 \tau'_{IDB} 
$.  This implies that when $\Delta \tau > \Delta \tau'_{\infty}$ the sum \nref{SumQ} diverges. This motivates the         imaginary distance bound \nref{IDBAdS} for any AdS$_D$ theory\footnote{ We thank the authors of \cite{OtherPaper} for explaining this point to us.}.  

One could still ask whether this sum reflects the right integration contour in the full path integral. In \cite{Held:2026huj} a different conclusion was reached, where the Lorentzian path integral and constrained instanton methods were used to argue that the answer was always finite in the AdS case, at least for the analytic continuations we are envisioning.

We note that the bound \nref{IDBAdS} has a simple holographic interpretation in terms of a dual boundary CFT.  The boundary values of scalar fields are CFT coupling constants, and the moduli space metric is the Zamolodchikov metric on the corresponding space of conformal field theories.  So \nref{IDBAdS} can be interpreted as bounding how far one can analytically continue in the space of complexified coupling constants.  Indeed, we already encountered bounds of this type in section \ref{sec:ftex}.

So far, we have discussed the case $D\geq 3$. The case of $D=2$ is special.  In JT gravity, 
there are wormholes with imaginary scalar fields, as we review in appendix \ref{WHJT}.  They can exist for any value of the boundary scalar field, and they have negative action. 
We will see in section \ref{DimReCh}, that these wormholes can arise from the near horizon region of extremal black holes. In that context, the flat space imaginary distance bound applies and we expect that the theory at these combined values of the imaginary scalar field and the chemical potential for the charge of the black hole should be problematic. However, one might wonder whether these imaginary values are problematic from a purely Nearly-AdS$_2$ point of view. One would expect that they should not be dominant in a UV complete model with no averaging over couplings. In the AdS$_2$ model with an average over complex couplings discussed in \cite{Garcia-Garcia:2020ttf} the wormhole is a perfectly fine solution.

\section{Wormholes from dimensional reduction}
\label{sec:dimensional}

Several special examples of wormhole solutions arise from Kaluza-Klein reductions. As is well known, Kaluza-Klein reduction on a circle gives gravity plus a massless scalar in one lower dimension. The wormhole solution involving this scalar is closely related to the Schwarzschild solution in the higher-dimensional space-time, and to the Saad-Shenker-Stanford wormhole \cite{Saad:2018bqo} based on that Schwarzschild solution. This example also allows us to relate the imaginary distance bound to the KSW criterion. 
We also consider a dimensional reduction of Einstein-Maxwell theory, which leads to a connection with the weak gravity conjecture.

\subsection{$S^1$ Reductions: The Schwarzschild Solution and SSS}

We consider a circle compactification from $D+1$ dimensions to $D$ dimensions.   Setting the Kaluza-Klein gauge field to zero, we can write  the metric ansatz 
\be  \la{KKRed}
ds^2_{D+1} = {\tilde r}^{ -2/(D-2) } g_{\mu \nu} dx^\mu dx^\nu + {\tilde r}^2 d\sigma^2    ~, ~~~~\sigma = \sigma + 2 \pi 
\ee 
which leads to the action  
\be \la{ActRed}
I =   { 1\over 16 \pi G_{N,D }} \int d^{D} x\sqrt{g} \left[-R + \half { 2 (D-1) \over (D-2) } ( \partial \nu)^2   \right] ~,~~~~\nu \equiv \log \tilde r 
\ee 
  Note that $\nu$ is non compact.

We can construct a Euclidean wormhole in $D$ dimensions using the scalar field $\nu$. The asymptotic values of $\nu$ on the two sides of the wormhole will differ by an imaginary amount  
\be 
\Delta \nu_{\infty} = i \pi 
\ee 
This is just the displacement \nref{Diff} described earlier, once one takes into account the $D$ dependent normalization factor in \nref{ActRed}.  If we take $\nu$ to be real at the wormhole throat, then this means $\tilde r$ will become purely imaginary as we approach the two asymptotic ends of the wormhole: $\tilde r\propto \pm i$. Thus the $\sigma$ circle will become 
Lorentzian at the two asymptotic ends of the wormhole. 
In fact the wormhole solution \nref{MeWH}, \nref{FaWH} is just the usual Schwarzschild solution in $D+1$ dimensions, albeit with a different choice of contour in the complex $r$-plane. This contour turns out to be similar to the contour chosen by \cite{Saad:2018bqo} for the double cone wormhole.\footnote{That the double-cone wormhole for the Schwarzschild black hole becomes the imaginary-scalar wormhole upon dimensional reduction is, in retrospect, more or less guaranteed by the $O(D)\times \R$ isometries of the Schwarzschild metric.}

\begin{figure}[h]
    \begin{center}
    \includegraphics[scale=.3]{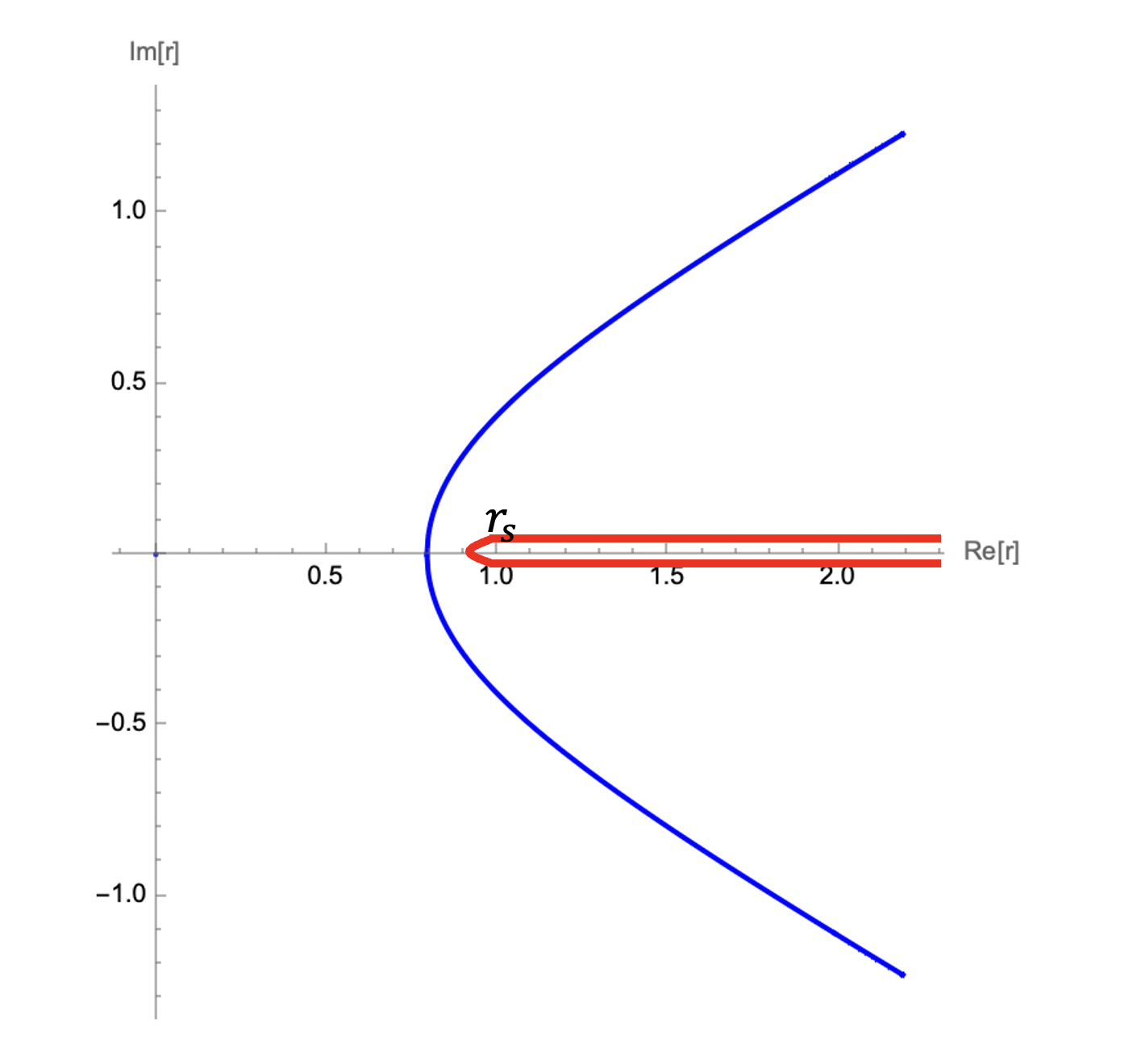}
    \end{center}
    \caption{ In red, we have the SSS \cite{Saad:2018bqo} contour for the double cone wormhole associated to the  Schwarzschild solution, which goes behind the horizon at $r_s=1$. In blue, we have plotted the contour that corresponds to real values of $a$ (for $D=5$ and $r_s=1$). We see that it is a simple deformation of the SSS contour.  }
    \label{PlotContour}
\end{figure}

To see this in more detail, let us start with the usual Schwarzschild black hole in $D+1$ dimensions:
 \be \la{SchSO} 
ds^2 = - f dt^2 + { dr^2 \over f} + r^2 d\Omega_{D-1}^2 ~,~~~~~~ f = 1 - {r_s^{D-2} \over r^{D-2} } 
\ee 
In lower dimensions, using \nref{KKRed}, this leads to the standard wormhole in \nref{MeWH}, \nref{FaWH}. Dimensionally reducing along the $t$ direction, and letting 
 \be 
 \tilde r^2 = e^{2 \nu} =- f \, , 
 \ee
 the $D$-dimensional metric is
 \be
 ds^2_D = (-f)^{  1 \over (D-2) } \left( { dr^2 \over f } + r^2 d\Omega_{D-1}^2 \right) =  { d a^2 \over F } + a^2 d\Omega_{D-1}^2 \, .
 \ee 
Here
 \be \la{rhoandr}
  a^{2 (D-2) } 
  = - f  r^{ 2(D-2) }   ~,~~~~~~{\rm i.e. } ~~~~ { r^{D-2} \over r_s^{D-2} } = { 1 \pm i \sqrt{ { a^{ 2 (D-2) } \over a_0^{2(D-2) } }-1} \over 2 }
  \ee with 
\be 
   a_0^{D-2 }= { 1 \over 2 } r_s^{D-2 } \, .
 \ee 
These formulas imply that real values of $a$ correspond generically to complex values of $r$. We see that the 
wormhole solution then corresponds to a particular section in the complex$-r$ plane of the standard Schwarzschild solution; this contour is plotted in Figure \ref{PlotContour}.\footnote{Note that 
if we started from the Euclidean black hole in $D+1$ dimensions, we would get a solution in $D$ dimensions where the scalar field is following a spacelike trajectory and the resulting $D$-dimensional metric has a singularity.} This is the contour in the complex$-r$ plane defined by
real $a\geq a_0$ in \nref{rhoandr}, including both branches of the square root; these two branches 
correspond to the two sides of the wormhole. 

This contour is very similar to the one chosen in the SSS double cone wormhole \cite{Saad:2018bqo}.  Indeed, many of the properties of imaginary scalar wormholes we have emphasized were already noted in \cite{Saad:2018bqo}. The fact that action is zero was important in \cite{Saad:2018bqo}. Moreover, the charge conjugate to $\nu$ translations is proportional to the mass of the black hole.  The size of the wormhole is related to the mass, and the fact that the classical solution has an action independent of the mass was also crucial in \cite{Saad:2018bqo}. Finally, we emphasize that $\nu$ is a non-compact scalar rather than an axion; the connection between the SSS wormhole and the imaginary scalar wormholes only becomes apparent when one considers general scalars, rather than just axions. 

In this particular case, the imaginary distance bound is given by $|{\rm Im}(\nu)| \leq \pi/2$. 
Note that if we choose a  flat space metric in \nref{KKRed} (\textit{i.e.} $\tilde r$ constant), this bound coincides with the Kontsevich-Segal-Witten criterion. 
That is because this bound limits the phase of the metric component in the $\sigma$ direction in \nref{KKRed} to be less than $\pi$.  In principle, the metric \nref{KKRed} might also have a phase coming from the $D$-dimensional part of the metric, but this can be removed by a phase rotation of those flat space coordinates by taking ${\tilde r }^{ -2/(D-2) } g_{\mu \nu} = \eta_{\mu \nu } $.

It is natural to ask whether the full wormhole also obeys the KSW criterion. It does not obey it in $D+1$ dimensions if we go along the blue contour in figure \nref{PlotContour}. However it is close to obeying it if we go along the red contour figure \nref{PlotContour}, although it is not obeyed exactly as we discuss in appendix \ref{KSWDC} (see also \cite{Chakravarty:2024bna}, where this point is made in a slightly different way). 

We note that, in this case, the $(D+1)$-dimensional theory makes sense for imaginary values of $\nu$ all the way up to imaginary distance bound $|{\rm Im}(\nu)| = \pi/2 $. This is in contrast with the dilaton of type IIB string theory and other examples we discuss later, where the problem with the theory arises before the IDB. In the higher-dimensional theory, the uplifted configuration is just the SSS double cone wormhole, 
which has a beautiful interpretation as computing the ramp in the spectral form factor \cite{Saad:2018bqo}.
In this ramp interpretation, the one loop prefactor was also important. Furthermore, \cite{Saad:2018bqo} restricted the ``charge'' (mass of the black hole) to be within some range. In contrast, our discussion here has been coarser, as we have only used the wormhole to set a bound on an imaginary deformation of a flat space compactification. For our discussion the very large size wormholes are the most relevant, where one-loop factors or other higher-derivative corrections are not important.  Similarly, the effects that are predicted by the IDB are very simple in this case: they are just particles that wrap the extra circle in the flat space compactification (the compactification with no wormhole). When $\nu \to i \pi/2$ these particles become tachyonic. 
However, the SSS \cite{Saad:2018bqo} discussion suggests that there is a stronger statement which holds charge by charge. Namely, in a complete theory,  for each wormhole  charge, there is some effect in the single sided theory that is of the same size or higher than the wormhole, as in the half wormhole discussion of \cite{Saad:2021rcu}. In this case, the prefactors are important.

Let us briefly  discuss some aspects of the mass of the KK modes as a function of the radius; these modes are often discussed in the context of the Swampland Distance Conjecture \cite{Ooguri:2006in,Grimm:2018ohb}. For pure gravity compactified on a single circle, the KK modes have mass
  \begin{align}
    m_{KK}^2 \sim n^2 \exp \left( - \sqrt{\frac{2(D-1)}{(D-2)}} \varphi\right) = n^2 \exp\left( - { 2 (D-1) \over ( D-2) } \nu \right) 
    \end{align}
    where $\varphi$ is a (canonically normalized) scalar field coming from the compactification. If $\varphi$ becomes large and real, corresponding to a large circle, we get light KK modes. However we can also take $\varphi$ to be complex. In particular, if we start at a large real $\varphi_0 $ and then add an imaginary displacement prescribed by the flat space IDB we find 
    \begin{align}
        m_{KK}^2 \sim n^2 \exp \left( - i \pi  \frac{D-1}{D-2} \right) \exp\left( - \sqrt{\frac{2(D-1)}{(D-2)}} \varphi_0\right) \, .
    \end{align}
    We see that in general we will move past the location where $m_{KK}^2$ is purely negative. So in this particular case, we eventually reach a displacement, which is smaller than the imaginary distance bound, where there is a tachyonic Kaluza-Klein mode. This seems consistent with the picture in figure \ref{FlatAllowed}, where a problem arises before the bound. 
    However, if we view the theory as a $(D+1)$-dimensional theory, there seems to be 
    no problem until we get to exactly the IDB. It would be interesting to analyze this problem further in order to find a closer connection between the Imaginary Distance Bound discussed in this paper and the Swampland Distance Conjecture.

\subsection{$S^1$ reductions with gauge fields: Charged black holes, and the Weak Gravity Conjecture}
\la{DimReCh}

Now let us consider a Kaluza-Klein reduction of a $(D+1)$-dimensional theory of gravity plus a Maxwell field,
 \begin{equation} \la{StDpO}
I_{D+1}
=
\frac{1}{16\pi G_{D+1}}
\int d^{D+1}x \sqrt{-  g}
\left[
 - R
+\frac14   F_{MN}  F^{MN}
\right], ~~~~ F = dA 
\end{equation}
 We write the metric ansatz 
\begin{equation}
d s^2_{D+1} \la{MetRedCh}
=
e^{-\frac{2\nu}{D-2}}\,ds_D^2
+
 e^{2\nu}d\sigma^2 ~,~~~~~~~~~ A = \alpha d\sigma 
\end{equation}
where $ds^2_D = g_{\mu \nu} dx^\mu dx^\nu$ is the $D-$dimensional Einstein metric. We have set to zero all the gauge fields in $D$ dimensions, we have only kept the holonomy $\alpha$ of the higher-dimensional gauge field on the circle. 
 
The dimensionally reduced action is
\be \la{MetMO}
I_{D}
=
\frac{1}{16\pi G_D}
\int d^Dx\,\sqrt{-g}
\left\{
-R
+
\half \hat \kappa^2 \left[ (\nabla\nu)^2
+
  e^{-2\nu}(\nabla \tilde \alpha)^2
\right] \right\}~,
\ee
with 
\be\hat \kappa^2 \equiv \frac{2 (D-1)}{D-2} ~,~~~~~~ \tilde \alpha \equiv  \alpha/\hat \kappa  \la{KappDe} .
\ee
Of course, the normalization of the $\nu$ field is the same as in \nref{ActRed}. We see that we now have a pair of scalar fields, $\nu$ and $\tilde \alpha$, which describe a moduli space whose target space metric is hyperbolic space with radius $\hat \kappa$.  
As in the previous section, we can reinterpret the Lorentzian black hole solution in $D+1$ dimensions as a wormhole in $D$ dimensions, where the scalars $(\nu, {\tilde \alpha})$ follow a geodesic motion through a complexified moduli space.  This is
quite similar to the discussion of the previous section, so we will not explore it further. 

To begin, note that if we make $\tilde \alpha $ imaginary, $\tilde \alpha = i t $, 
the moduli space metric becomes a Lorentzian AdS$_2$ with radius $\hat \kappa $ \nref{KappDe}. The distance between the real section, at $t=0$, and the Poincaré horizon is $\hat \kappa {\pi \over 2 }$, which is precisely the imaginary distance bound \nref{IDB}. The Poincaré horizon is the region with 
\be \la{PoinCH}
 e^{\nu }  ,  \pm t  \to \infty 
 \ee 
while the Poincaré patch is the region with $e^\nu$ positive, as in Figure \ref{AdS2Instanton}.  The same distance, $\hat \kappa {\pi \over 2 }$, separates the real slice at $t=0$ with the lines (see figure \ref{AdS2Instanton})
 \be \la{ExtBou}
 e^{\nu }  \pm t  = e^\nu \pm i \tilde \alpha =0 ~.
 \ee 
The action of a charged particle wrapping the $\sigma $ direction in \nref{MetRedCh} is 
\begin{equation}
    S = -m \int ds \sqrt{g_{\sigma \sigma}} + i q \int A = - 2 \pi \left( m \tilde r - i q  \hat \kappa \tilde \alpha \right) = - 2\pi m ( e^\nu - { \hat \kappa q \over m } t ) 
\end{equation}
On the lines \nref{ExtBou}, this vanishes when $m = \hat \kappa q$. This charge-to-mass ratio 
coincides with the charge-to-mass ratio of extremal charged black holes in $D+1$ dimensions. 
Any  particle with $m <  \hat \kappa q $ would lead to large corrections when we cross the lines \nref{ExtBou}.  The existence of such a particle is precisely what is ensured by the weak gravity conjecture \cite{Arkani-Hamed:2006emk} (see also \cite{Harlow:2022ich} for a review of the many forms of the conjecture), which for the Lagrangian in \nref{StDpO} requires a particle satisfying
\begin{align}
    \frac{|q|}{m} \, > \, \sqrt{\frac{D-2}{2(D-1)}} \, ,
\end{align}
We conclude that, in this case, the weak gravity conjecture coincides with the IDB bound, phrased as in  \nref{NDWC}.

In principle, we could stop the discussion here. However, it is  instructive to discuss wormholes a bit more. First note that we can rewrite the solution \nref{MeWH} in conformal coordinates as 
 \be \la{ConfWH}
 ds^2_D = a^2 \left( { d \rho^2 \over (D-2)^2 }  + d\Omega_{D-1}^2 \right) ~,~~~~~~~  { a^{D-2} \over  a_0^{D-2}} = \cosh\rho = { 1 \over \cos \tilde \tau }~,~~~~
 \ee 
 where we have picked a convenient normalization for the radial coordinate. For large $\rho$, $a \propto e^{ \rho/(D-2) } $ and the metric is asymptotically flat, just as in \nref{MeWH}.
 
We will consider the simplest scalar field wormhole in $D$ dimensions that corresponds to the solutions discussed previously, and describes a geodesic in the AdS$_2$ moduli space:
\be 
 e^{ \nu } = e^\gamma { 1 \over \cos \tilde \tau } ~,~~~~~~ i \tilde \alpha = e^\gamma \tan \tilde \tau 
\ee 
where $\tilde \tau = \tau/\hat \kappa$, and $\tau$ is the proper time coordinate along the geodesic. 

We can uplift this to $D+1$ dimensions by plugging \nref{ConfWH} into \nref{MetRedCh}, which gives 
\be \la{BRSol}
ds_{D+1}^2 = r_e^2 \left[  { 1 \over (D-2)^2 } \left( \cosh \rho^2 d\hat \sigma^2 + d\rho^2 
\right) + d\Omega_{D-1}^2  \right] ~,~~~~~ r_e^2 = a_0^2 e^{ - 2\gamma/(D-2) } 
\ee 
where $\hat \sigma = \sigma / (D-2) $. We can easily see that this is precisely AdS$_2 \times S^{D-1}$. Note that AdS$_2$ is in global coordinates; the two boundaries of global AdS$_2$ are the two ends of the wormhole. 
Curiously, we note that the wormhole is asymptotically flat in $D$ dimensions, but is not asymptotically flat in $D+1$ dimensions. This is because the scalar $\nu$ is going to infinity at the endpoints of the wormhole, which happens precisely because the wormhole trajectory is the exact length to extend between the past and future ``Poincar\'e horizons'' on moduli space; this is denoted in figure 
\nref{Geodesics}(b).  

The general picture we arrive at is the following: wormhole solutions to \nref{MetMO} uplift to wormhole solutions of Einstein-Maxwell theory that are AdS$_2 \times S^{D-1}$. This entire class of wormholes is destabilized by superextremal particles. These particles wrap the Kaluza-Klein circle, and they appear as instantons in the $D$-dimensional theory\footnote{See also appendix \ref{app:axions} for a discussion of the axion weak gravity conjecture, which concerns instantons and can be related to the WGC by dimensional reduction \cite{Heidenreich:2015nta}.}. So the weak gravity conjecture ensures that these AdS$_2 \times S^{D-1}$ wormhole solutions to Einstein-Maxwell theory will break down. Of course, this is not a surprise; the weak gravity conjecture was designed to render extremal black holes unstable \cite{Arkani-Hamed:2006emk}.  Note that there are also other important effects, such as quantum corrections; see \cite{Mertens:2022irh}.
  
\subsubsection{Constructing a more regular wormhole solution } 
\la{ChRegWH}

The previous wormhole solution \nref{BRSol} was somewhat unusual because $\nu$ diverged at the ends of the wormhole. This was because the scalar moduli space was an AdS$_2$ with radius \nref{KappDe}, which turned out to be precisely the value for which 
the geodesic distance between the two horizons is the distance needed to support a wormhole. One can construct a more regular wormhole solution by adding additional fields which will roll as well.  So we would have a wormhole solution depicted in figure \ref{Geodesics}(c) rather than figure \ref{Geodesics}(b). 
 
We add the additional field $\chi$ to the Einstein-Maxwell theory,
\nref{StDpO}. After reducing to $D$ dimensions we get 
\begin{equation} \la{MetMOs}
S_{\rm scalar}
=
\frac{1}{16\pi G_D}
\int d^Dx\,\sqrt{-g}
\left\{
R
-
\half \hat \kappa^2 \left[(\nabla\nu)^2
+
  e^{-2\nu}(\nabla\tilde \alpha )^2 +  (\nabla \chi)^2 \right] 
\right\} ~,~~~~~~ 
\end{equation}
where we have normalized $\chi$ to include a factor of $\hat \kappa$ \nref{KappDe} for convenience. 

We can now construct a wormhole solution by 
considering a more general geodesic in the three-dimensional scalar field moduli space with coordinates $(\nu, {\tilde \alpha}, \chi)$. Such a geodesic will simultaneously be
a geodesic in the complexified $H_2$ directions, as well as a straight line in the imaginary $\chi$ direction. The geodesic in $H_2$ direction will now have a shorter imaginary distance. More explicitly, the metric in $D$ dimensions is as above, but the moduli will take the form  
\be \la{GeoReCh}
e^{ \nu } = e^\gamma { 1 \over \cos\left(   \tilde \tau   \cos \epsilon \right)  } ~,~~~~~~~ \tilde \alpha  = i e^\gamma \tan \left( \tilde \tau  \cos \epsilon \right) ~,~~~~~~~~ \chi = i \tilde \tau   \sin \epsilon 
~,~~~~\tilde \tau  \equiv { \tau \over \hat \kappa}  \ee 
where the parameter $\epsilon $ determines the relative speed of the geodesic in the $H_2$ directions compared to the
$\chi $ direction. 

For $\epsilon > 0$, we now find that the metric in $D+1$ dimensions is asymptotically flat and is a type of traversable wormhole. The geometry is similar to that of an extremal black hole, and also to the traversable wormholes constructed from quantum effects \cite{Maldacena:2018lmt}. Here we get negative energy from the complex $\chi$ field, but the solution is very similar. Of course, the fact that we have a complex $\chi$ field implies that the solution is not a solution for a unitary theory. In Appendix \ref{AppCharged}, we give more details of this solution. Here, we only wanted to point out that the addition of a small imaginary displacement of another field is enough to turn this into an asymptotically flat traversable wormhole in $D+1$ dimensions. 

As an additional comment, note that it is straightforward to consider a situation with many Maxwell fields, with a kinetic term proportional to 
\be 
\int G_{IJ} F^I_{\mu \nu} F^{J \mu \nu} ~,~~~~~~ \to ~~~~~~~G_{IJ} \nabla \alpha^I \nabla \alpha^J 
\ee 
where $I=1, \cdots , K $ runs of the $K$ Maxwell fields. 
We have also included the moduli space metric $G$. Now the wormhole involves a straight line in this $\alpha$ space. The imaginary distance bound carves out a sphere in this $K$-dimensional space. Demanding the existence of charged particles that appear before we get to the surface of the sphere carved out by the imaginary distance bound, we get the familiar convex hull condition for the WGC with multiple gauge fields \cite{Cheung:2014vva}. 

A final connection with the WGC is the following.   Higher-derivative terms can correct the extremality bound, which allows extremal black holes to be slightly self-repulsive and satisfy the WGC \cite{Kats:2006xp}. The corrections to the wormhole action were computed in \cite{Andriolo:2020lul}. It should be the case that with a particular sign of higher-derivative corrections, the fixed-charge wormholes can similarly fragment into smaller fixed-charge wormholes. It would be interesting to see if the corrections to the black hole mass in \cite{Kats:2006xp} can be related to the corrections to the wormhole action in \cite{Andriolo:2020lul} through dimensional reduction.

\subsection{More General Reductions and the KSW Condition}

In this section, we explore the relationship between our imaginary distance bound \nref{IDB} and the Kontsevich-Segal-Witten (KSW) criterion. This criterion, introduced by Kontsevich and Segal \cite{Kontsevich:2021dmb} as a condition on field theories, was proposed by Witten \cite{Witten:2021nzp}
as a constraint on which complex metrics 
can be regarded as legitimate saddle points of a gravitational path integral.
For a complex metric $g_{\mu\nu}$ with eigenvalues $\lambda_i$, the condition is that 
\begin{align}
    \label{eq:KSW_cond}
    \sum_{i=1}^D |\mathrm{Arg}(
    \lambda_i)| < \pi \, ,
\end{align}
Here the principal branch of the Argument is chosen. This condition follows from the requirement that, when a canonically coupled $p$-form field is placed in this background, all of the resulting Gaussian fluctuations are well behaved (so fluctuations are suppressed in the path integral, rather than enhanced).  This is a strong requirement -- for example, a Lorentzian signature space-time saturates this bound -- but one which guarantees that perturbative field theory is well-defined.  An interesting complex geometry that obeys it was discussed in \cite{BenettiGenolini:2026raa}.

Here we will consider more general torus compactification and concentrate on the part of the moduli space that comes from the geometry of the internal torus. We will argue that in this particular case the imaginary distance bound \nref{IDB} coincides with the KSW criterion.

To make this explicit, consider a compactification of a $(D+N)$-dimensional space-time down to $D$ dimensions on a $N-$dimensional torus $T^N$.  We will write the full $(D+N)$-dimensional metric as
\begin{align}
    ds^2 = e^{-\frac{N}{D-2} u} ds^2_{D} + e^{u} m_{ij} dx^i dx^j
\end{align}
where $x^i$ are the internal directions and $m_{ij}$ is a symmetric matrix with unit determinant.  Both $u$ and $m_{ij}$ are functions of the $D$-dimensional coordinates, and will appear in the low-energy $D$-dimensional theory as scalar moduli $\varphi^a=(u,m_{ij})$ that parameterize the volume and shape of the torus, respectively.
The   dimensionally reduced action is the usual Einstein-Hilbert term in $D$ dimensions, 
\begin{equation}
S = \frac{1}{2\kappa_{D}^2}\int d^{D}x \sqrt{g}\left(-R+\frac{1}{2} h_{ab} \partial_\mu \varphi^a \partial_\nu\varphi^b\right)
\end{equation}
where $g$ is the determinant of the $D$-dimensional metric $ds_D^2$, and $R$ is its Ricci scalar. 
The  metric on moduli space is 
\begin{equation}
ds^2_{\cal M}\equiv h_{ab}d\varphi^a d\varphi^b = \frac{2(D+N-2)}{N(D-2)}du^2 + \frac{1}{2} \Tr\left(m^{-1} dm\right)^2
\end{equation}
The metric for the shape modulus
is built out of the left-invariant one-form $m^{-1} dm$, where $m=m_{ij}$.\footnote{The space of symmetric matrices $m$ can also be represented as the coset $SL(N,\R)/SO(N,\R)$.  Any symmetric matrix $m$ with unit determinant can be written as $m=\gamma \gamma^T$ for some $\gamma\in SL(N,\R)$.  This representation is invariant under right multiplication $\gamma\to \gamma R$ by a matrix $R\in SO(N,\R)$, allowing us to identify the space of symmetric matrices as a coset. 
The coset has a natural $SL(N,\R)$ action by left multiplication, which takes $\gamma\to L \gamma$ and $m\to L m L^T$ with $L\in SL(N,\R)$. 
From the point of view of the low-energy theory, the matrix $m_{ij}$ corresponds to $N-1$ ``dilatons'' (\textit{i.e.} D-dimensional scalar fields) and $\frac{N(N-1)}{2}$ ``axions'' (\textit{i.e.} D-dimensional pseudo-scalar fields). } 

We now imagine complexifying this moduli space, by letting the  $\varphi^a=(u,m_{ij})$ be complex.  We still require that $m$ is symmetric with unit determinant, as its determinant can be absorbed into $u$.\footnote{Again, the space of symmetric complex matrices with unit determinant is the coset $SL(N,\C)/SO(N,\C)$.}   
The metric $ds^2_{\cal M}$ on moduli space is the same as above, but now can take complex values.  Similarly, the metric $e^u m_{ij} dx^i dx^j$ on the torus is now complex as well.
In a wormhole solution the $\varphi^a$ follow a Lorentzian geodesic through this complexified moduli space.  Since the moduli space metric factorizes into a volume mode and the shape modes, this is just a geodesic simultaneously in these two parts of moduli space.  We wish to find the longest (timelike) geodesic possible which still obeys KSW.  

In order to analyze the KSW criterion, it is convenient to work with the full internal metric $G_{ij} = e^u m_{ij}$, in terms of which the moduli space metric is 
\begin{equation}\label{modmec}
ds^2_{\cal M} = \frac{1}{2} \Tr \left(G^{-1} dG\right)^2 +\frac{1}{2(D-2)} \left(\Tr \left(G^{-1}dG\right)\right)^2
\end{equation}
There is a natural $GL(N,\C)$ action on the space of complex symmetric matrices, which takes $G\to \gamma G\gamma^T$ for $\gamma\in GL(N,\C)$.  The metric $ds^2_{\cal M}$ is invariant under this action.
We now consider a timelike geodesic $G(t)$ through moduli space.
In fact, a short computation shows that the geodesic equation is solved by $G = \gamma e^{ i P t} \gamma^T$ for some $\gamma \in GL(N,\C)$ and complex symmetric $P$.  Moreover, the timelike distance will be maximized when $P$ is real.  Thus $P$ is diagonalizable, so we can write 
$G(t) = \gamma D(t) \gamma^T$ for some diagonal $D(t)$ and a fixed $\gamma \in GL(N,\C)$. Let us write the eigenvalues of $G$ as $e^{i \theta_i}$, so the metric is 
\begin{equation}
\label{modmectheta}
ds^2_{\cal M} = -\frac{1}{2} \sum_i d\theta_i^2 - \frac{1}{2(D-2)} \left(\sum_{i} d\theta_i\right)^2
\end{equation}

We wish to construct the longest possible timelike geodesic in moduli space which is consistent with the KSW criterion.  This will be accomplished by taking the $\theta_i$ to be real.  The KSW criterion for the internal part of the metric is 
\begin{equation} \la{PhasCons}
\sum_i |\theta_i|<\pi    
\end{equation}

We now consider the imaginary geodesic distance $\tau$ between a real metric and the point
which obeys \nref{PhasCons}. From \nref{modmectheta}, this geodesic distance is
 \begin{equation}
  \tau^2 = \frac{1}{2} \sum_i  \theta_i^2 + \frac{1}{2(D-2)} \left(\sum_{i}  \theta_i\right)^2
\end{equation}  
Using the inequalities
\begin{equation}
\sum_i  \theta_i^2 \le \left(\sum_i | \theta_i|\right)^2,~~~~~\left(\sum_i   \theta_i\right)^2 \le \left(\sum_i |  \theta_i|\right)^2
\end{equation}
along with the KSW criterion \nref{PhasCons} we have 
\begin{equation} \la{ADis}
  \tau^2 <  \pi^2 \left(\frac{1}{2} + \frac{1}{2(D-2)}\right) = { \pi^2 \over 4 }  \frac{2(D-1)}{D-2}
\end{equation}
This is {\it exactly} the flat space imaginary distance bound \nref{IDB}.

Note that in applying the KSW bound above, we have ignored the contributions from the $D$-dimensional metric. 
But no matter what this metric is (either real or complex) it can only make the bound stricter.

We emphasize that a compactification will also have non-geometric moduli. In that case, the Imaginary Distance Bound \nref{IDB} can be viewed as a generalization of the KSW criterion for a complexified compactification.

We note that our computation is similar to, but technically independent from, the analysis performed in \cite{Loges:2022nuw}. In that paper, the authors analyzed the perturbative stability of the fields that comprise the solution, including the axion, dilaton, and gravity. The KSW condition, on the other hand, concerns the stability of fields that may not even be part of the theory, and it does not include modes of the metric itself. It would be interesting to understand what exactly is the full set of criteria for perturbative stability. For instance, when are complex background form fields allowed in general? Is there a simple criterion like the KSW condition for stability against metric fluctuations, or is the Picard-Lefschetz analysis of \cite{Loges:2022nuw} required? As a first step, it would be interesting to know if there is a minimal 
generalization of the KSW condition which is manifestly invariant under dualities that exchange the metric and other fields upon compactification.

\subsection{Branes in flat space and AdS wormholes }

There is a connection between the flat space and AdS wormholes. The idea is very simple; we view AdS$_D$ as arising from an extremal black brane in higher dimensions.  Then the AdS$_D$ wormholes with flat slicing, as in \nref{PoinAct}, can be viewed as a particular type of flat space wormholes in $p+1$ dimensions.  

We start with a theory of gravity in $D+p$ dimensions with  a $D-1$ form potential and a scalar field $\varphi $ 
\be 
I =   { 1 \over 16 \pi G_{N, D+p} } \int d x^{D+p} \sqrt{G} \left[- R  + { 1 \over 2 D!} F^2_D  + \half (\nabla \varphi)^2 \right] ~,~~~~F_D = d A_{D-1}
\ee 
We now take the KK ansatz 
\bea 
ds^2_{D+p} &=& e^{ - {2   \over (p-1) }  \sigma } g_{\mu \nu } dx^\mu dx^\nu + e^{ 2   \sigma \over (D-1) } d\vec y^2 ~,~~~~~~~~ A = \alpha  dy^1 \cdots dy^{D-1}
\eea 
where $g_{\mu \nu}$ is the metric in $p+1$ dimensions and $\vec y$ a $(D-1)$-dimensional torus. 
We then find 
\bea \la{DimRedF}
I &=&   { V_T \over 16 \pi G_N} \int d^{p+1} x \sqrt{g} \left[- R^{(p)} + \half { 2   (D+p-2) \over (D-1) ( p-1) } (\nabla  \sigma)^2 + \half e^{ - 2    \sigma } (\nabla \alpha )^2 + \half (\nabla \varphi)^2 \right]  ~~~~~~~~~
\eea 
The extremal black $(D-2)$-brane solution has the form, see \textit{e.g.} \cite{Stelle:1996tz}, 
\be \la{EBrane}
ds^2= H^{- {2 \over D-1}} d\vec y^2 + H^{ 2 \over p-1 } (dr^2 + r^2 d\Omega_p^2) ~,~~~~~H = 1 + { Q \over r^{p-1} } ~,~~~~ \alpha  \propto  i H^{-1} 
\ee 
As a first check, we would like to reproduce this extremal brane solution from scalar field plus gravity formulas in $\hat D = p+1$ dimensions. 
The extremal brane solution \nref{EBrane} has the following form in $\hat D$ dimensions 
\be \la{LLTra}
e^{   \sigma } = H^{- 1}    ~,~~~~~ a^2 = r^2  ~,~~~~~~~\alpha \propto i H^{-1} 
\ee 
 This is a light-like trajectory in the moduli space, where the $\hat D$-dimensional metric is just flat space.  The imaginary value of $\alpha$ can be viewed as a real chemical potential for the brane charge. 

We now want to consider a wormhole-like trajectory that would look like \nref{LLTra} far away. The moduli space has an $H_2$ part spanned by the coordinates $\sigma$ and $\alpha$ whose radius is $\kappa^2 = {  2 ( D+p -2) \over (D-1) (p-1) } $. In addition we have the scalar field $\varphi$. We want a timelike trajectory both in the $H_2$ part as well as for the $\varphi$ part. 
The natural length of a trajectory that explores a large range of values of $\sigma$ would be one that fits in the Poincar\'e patch, such as the one in figure \ref{Geodesics}(b). Such a geodesic has a total imaginary length squared 
\be 
 \tau^2 = \pi^2  {  2 ( D+p -2) \over (D-1) (p-1) } 
 \ee 
 This plus the imaginary length squared in the $\varphi$ direction should sum up to the appropriate value for a wormhole in $\hat D =p+1$ dimensions  
\be \la{FindDi}
\pi^2 { 2 p \over p-1} = \pi^2  {  2 ( D+p -2) \over (D-1) (p-1) }  +  \left|\Delta {\rm Im} \varphi\right|^2 ~~~\to ~~~~\left|\Delta {\rm Im} \varphi\right|^2 = -\pi^2 { 2 (D-2 ) \over D-1} 
\ee 
which is indeed what we had in \nref{PoinAct}. This is the large $a_0$ limit of the total field displacement for the AdS$_D$ wormhole \nref{AdSSol} and it is the lower value in \nref{RangeDel}.   This shows that this lower value of field displacement in AdS$_D$ corresponds to an instability, at least for the case where we have the AdS$_D$ embedded in higher dimensions with a flat torus boundary. 

Notice that $p$ has disappeared from $\Delta \varphi$, so we could have chosen various values of $p\geq 2$. This is indirect evidence that this represents a limit more intrinsic to AdS$_D$ than the particular embedding.

\section{Wormholes in UV complete theories}
\label{sec:UV}

We have argued that the existence of wormholes gives an absolute upper bound on the imaginary distance in moduli space between two well-defined field theories.  In some cases  there may be a problem earlier, meaning before we reach the full wormhole distance. In this section we will discuss a few such examples.

\subsection{Wormholes in type IIB string theory}
\label{sec:WHinST}

Perhaps the simplest and most canonical example is type IIB string theory in AdS$_5 \times S^5$, dual to ${\cal N}=4$ Yang-Mills in the large $N$ limit. The moduli space metric is
\begin{align}
    \label{eq:modAdS}
    ds_\mathcal{M}^2 \ = \ d \phi^2 + e^{2 \phi} d C_0^2 \, .
\end{align}
and it is the same in $D=10$ or $D=5$ dimensions. 
The dilaton $\phi$ and axion $C_0$ are dual to the Yang-Mills coupling and the theta angle, respectively. 
The theory with only axion, dilaton, and graviton is a consistent truncation of IIB supergravity in 10 dimensions, so that we can restrict to these fields when we search for  gravity solutions.  
In the ten-dimensional language, the axion is a Ramond-Ramond 0-form and the dilaton is related to the string coupling by $g_s = e^\phi$. Let us define $Z = e^{-\phi}$ and $T = -i C_0$. Then the full set of solutions for $D > 2$,
\begin{align}
    \label{eq:scalareqgeneral}
    e^{-\phi} = Z \ &= \ e^{\gamma} { 1 \over \cos\left( \tau + c_0 \right) } \, , \\  -i C_0 = T \ &= \ e^{\gamma} \tan \left( \tau + c_0 \right) + c_1 \, , \nn
\end{align}
with integration constants $\gamma$, $c_0$, and $c_1$. A fourth integration constant controls the geometric size of the wormhole,  which sets the axion charge.   

\paragraph{Supergravity} 

We first explore the field range purely from the supergravity point of view. As we make $C_0 = i T $ imaginary we get a moduli space which is  AdS$_2$. In that space, we might want to assign some special significance to the region with $Z= e^{-\phi } >0$. Just from the point of view of gravity we can consider a bigger region. We go to global AdS$_2 $ coordinates by defining 
\begin{equation}\label{cov}
e^{-\phi } \equiv Z =\f{\cos y}{\sin y+ \cos t},~~~~~T=\frac{\sin t}{\sin y+ \cos t}
\end{equation}
This is just a field redefinition and the moduli space metric becomes 
\begin{equation}
\label{eq:global}
ds^2_\M 
= \frac{dy^2-dt^2}{\cos^2 y} ~,~~~~~~~~ y \in \left[ - { \pi \over 2 } , {\pi \over 2} \right]
\end{equation}
A timelike geodesic  oscillates around $y=0$. A simple example is the geodesic with $y=0$ and any $t$, which is   
\be
\label{eq:geodesic}
Z ={ 1 \over \cos t },~~~~~T=\tan t
\ee
in the original coordinates. In this coordinate system there is no problem when $Z$ becomes negative.

\paragraph{String theory and string instantons}
 
String theory is a different case. We see that when $Z$ flips sign, for instance at $t = \pi/2$, the string coupling also flips sign. This will definitely be a problem in string theory. For this reason, searches for wormholes have tried to construct them so that they lie within the ``Poincar\'e patch'', with $Z> 0 $, \textit{e.g.} \cite{Arkani-Hamed:2007cpn}. 

In addition, there are new states that can cause problems for wormholes. 
For example,   we have instantons, or $D(-1)$ branes, with action  
\begin{align}
    I_\pm = { 1\over g_s }  \pm i  C_0  = Z\pm T \la{ActInsZ}
\end{align}
with plus/minus corresponding to the instanton/anti-instanton. 
These vanish along two null lines in AdS$_2$, see dotted lines in figure \ref{AdS2Instanton}. These are
{\rm distinct} from the Poincar\'e horizons at $Z=\pm t =\infty $, which are the solid lines in figure \nref{AdS2Instanton}. String theory is singular on both. Geodesics which are reflection symmetric around the real slice with $T=0$ cross both at the same time. 

\paragraph{Higher-derivative corrections} 

In order to see more vividly how trouble arises, it is useful to focus on higher-derivative corrections. We define 
\be \la{TausHere}
{\cal T} = {\cal T}_1 + i {\cal T}_2 = C_0 + i  { 1 \over g_s} = i (Z + T ) ~,~~~~~~ \overline{ \cal T }  = C_0 - { i } { 1 \over g_s } = - i (Z - T) 
\ee 
Of course, when $C_0$ is imaginary, these are not complex conjugates. 
 Green and Gutperle \cite{Green:1997tv} argued that the $R^4$ correction goes like 
\begin{align}
    I_{R^4} \sim \, \mathcal E_{3/2}({\cal T }, \overline{ \cal T }) \,  R^4
\end{align}
Note that we get an $R^4$ term because the instantons break half of the supersymmetries. Each broken supersymmetry leads to an integration over a superspace variable and contributes to a half-derivative. So sixteen broken supersymmetries lead to the eight derivatives in the $R^4$ term. A wormhole would break all 32 supersymmetries and should arise from averaging over terms with sixteen derivatives.  

The function $\mathcal E_{3/2}$ is an Eisenstein series 
\begin{align}
    \label{eq:SL2Eisenstein}
   \mathcal E_{3/2}({\cal T }, \overline{ \cal T })\ = \ \sum_{(m, n) \neq (0,0)} \left[ \frac{{\cal T}_2}{( m + n {\cal T})( m + n \overline{\cal T})} \right]^{3/2}
\end{align}
We see that when $I_+$ or $I_-$ \nref{ActInsZ} are zero, then either ${\cal T}$ or $\overline{\cal T} $ are zero, and \nref{eq:SL2Eisenstein} is infinite, due to the part of the sum with $m=0$. The fact that this divergence comes from the sum over instantons is more clear to see if we express \nref{eq:SL2Eisenstein} as
\begin{align} \la{InstSu}
    \mathcal E_{3/2}({\cal T }, \overline{ \cal T }) = 2 \zeta(3) {\cal T}_2^{3/2} + \frac{2 \pi^2}{3} {\cal T}_2^{-1/2} + 8 \pi {\cal T}_2^{1/2} \sum_{{\tilde m} \neq 0, \, n \geq 1} \left| \frac{\tilde m}{n} \right| e^{2 \pi i {\tilde m} n {\cal T}_1} K_1(2 \pi |{\tilde m}n| {{\cal T}}_2) \, .
\end{align}
In this expansion, the first two terms are the perturbative tree-level \cite{Gross:1986iv} and one-loop corrections \cite{Green:1981ya}, while the sum takes the non-perturbative $D(-1)$ instanton effects into account. When $i{\cal T}_1 = \pm {\cal T}_2 $ we find that the sum over $\tilde m$ in \nref{InstSu} diverges. Since $\tilde m$ is the Poisson resummed version of  the variable $m$ in \nref{eq:SL2Eisenstein}, the large $\tilde m $ divergence is related to the $m=0$ one in \nref{eq:SL2Eisenstein}.

We then conclude that if we look at a configuration with fixed size in Planck units, then the higher derivative corrections become dominant as we cross the boundaries of analyticity at ${\cal T} =0$ or $\overline{\cal T} =0$. These are the boundaries at $Z=\pm T $ denoted by dotted lines in figure \ref{AdS2Instanton}.

\subsection{Torus compactifications}

There is an interesting class of wormholes, first discussed in \cite{Arkani-Hamed:2007cpn}, which follow geodesics that do not cross a horizon or a line $Z=\pm T$ (see figure \ref{AdS2Instanton}).  This is accomplished using multiple scalar fields -- by setting all fields equal, the total distance $\tau$ in moduli space can reach the wormhole distance without any one field moving too much.  

Let us  start with the ten-dimensional action
\begin{align}
    I = \frac{1}{16 \pi G} \int d^{10}x \sqrt{-G} e^{-2\Phi_{10}} (- R_{10} + \frac{1}{12} H^2 - 4 \partial^\mu \Phi_{10} \partial_\mu \Phi_{10} )
\end{align}
Let us first dimensionally reduce on 
  $T^2$. This leads to four extra moduli, $G_{9,9}$, $G_{9,10}$, $G_{10,10}$ and $B_{9,10}$. We will ignore the vector fields and external components of $B$, which can be consistently truncated. 
We go to Einstein frame $g_{\mu \nu} \to e^{-\frac{2}{3} (\Phi - \Phi_0)}g_{\mu \nu}$, and define
\begin{align}
    \rho = B_{9,10} + i \sqrt{ \det{}  G_{mn}}  \, , \qquad 
    ds^2 = \frac{\rho_2}{\sigma_2} |dx_{9} + \sigma dx_{10}|^2 ~,~~~\Phi_8 = \Phi_{10} - 
    \half \log \rho_2. 
\end{align}
where $\sigma = \sigma_1 + i \sigma_2$. 
We get 
\begin{align}
        I &= 
       \frac{1}{16 \pi G_8}  \int d^{8}x \sqrt{-g}  \Big[ -R + \frac{(\partial \sigma_1)^2 + (\partial \sigma_2)^2}{2 \sigma_2^2} + \frac{(\partial \rho_1)^2 + (\partial \rho_2)^2}{2 \rho_2^2} + \half (\partial \Phi_8)^2\Big]  
\end{align}
We set the dilaton $\Phi_8=0$. Then taking a 
diagonal path $\rho = \sigma$, we find an effective hyperbolic space of radius $\sqrt{2}$. In eight-dimensional flat space the radius would need to be $\sqrt{7/3}$ to fit a wormhole \nref{IDB}. So we find that in this system the geodesics on moduli space do cross a horizon or a vanishing instanton action line, see figure \ref{AdS2Instanton}. 

\subsubsection{$T^4$ compactification}

\begin{figure}[h]
		\centering
		\includegraphics[width = 10.0cm]{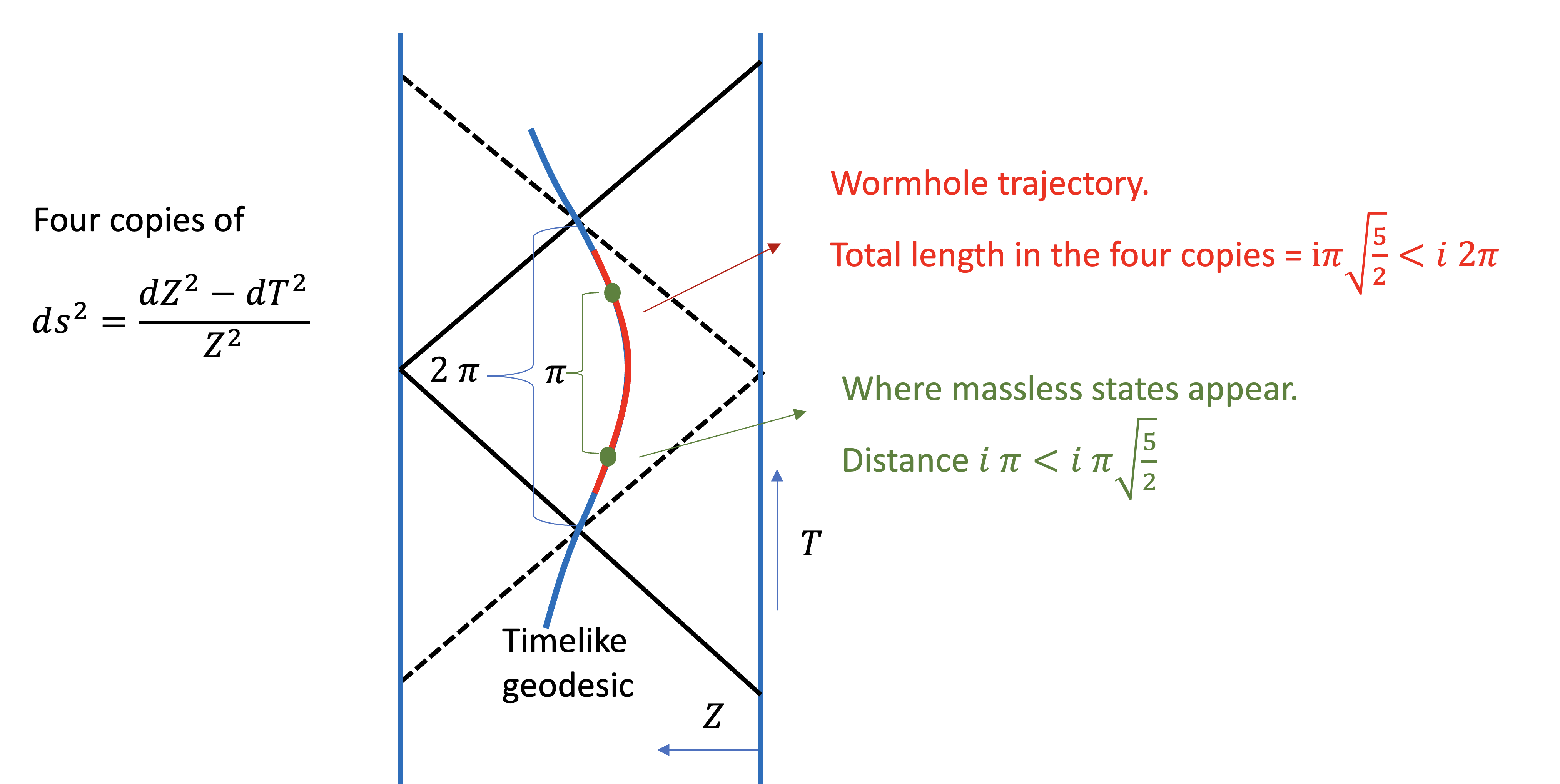}
    \caption{The wormhole solution in \cite{Arkani-Hamed:2007cpn} is based on a moduli space that has four copies of a radius-one hyperbolic space. We consider a ``diagonal'' trajectory that is the same on the four copies. The wormhole trajectory for  $D=6$ wormhole has total length $\pi\sqrt{5/2}$ and its trajectory on one of the copies is depicted here in red. It sits within the region of positive $Z$ and also within the region where $Z \pm T > 0$, where the simplest instanton effects are suppressed. Nevertheless, the theory breaks down at the green dots, where some momentum and winding states become tachyonic, or equivalently where some instanton action becomes negative. 
    }
    \label{T4WH}
\end{figure}

It is easy to generalize the above construction by compactifying on more tori -- these \textit{will} satisfy the length criterion of \cite{Arkani-Hamed:2007cpn}: wormholes can be formed without crossing horizons or vanishing instanton actions in moduli space. Viewing $T^4 = T^2 \times T^2 $ and focusing on the moduli of just the two two-tori, we find a second set of moduli $\sigma'$ and $\rho'$. So now we have four hyperbolic spaces of radius one. We see that $\pi \sqrt{ 5/2} $ (the wormhole length in $D=6)$  is smaller than $2\pi $ (the length of a diagonal geodesic that just fits between the two horizons or two zero action instanton lines), see figure \ref{T4WH}. 
For this reason, these wormholes appear to be completely safe from any correction. 
In this section, we explore this question in detail and show that there is a correction that appears {\it before} the wormhole distance, 
see figure \ref{T4WH}. 

{\bf Worldsheet instantons} 
Strings that are entirely wrapped on the compact dimensions appear as instantons in the six-dimensional theory. These wrapped strings will have 8 winding numbers -- a four-vector $w^i$ and $\tilde w^i$ for each of the dimensions of the worldsheet, $\xi_1$ and $\xi_2$. The compactified coordinates $x^i$ are then
\be 
x^i = w^i \xi_1 + \tilde w^i \xi_2 ~,~~~~~~ \xi_i = \xi_i + 2\pi ~
\ee 
The Nambu-Goto part of the action is 
\be \la{Nambu}
I_{NG} = { {\rm Area} \over 2 \pi \alpha'} =  2 \pi  \sqrt{ (w. w) (\tilde w . \tilde w )- ( w. \tilde w)^2   } 
\ee 
where  $a . b = a^i G_{ij} b^j$, with $G$ the metric on the $T^4$. This $T^4$ is given by the product of two two-tori with parameters obeying $\sigma = \sigma' = \rho = \rho'$. 
This means that the full action, including \nref{Nambu} and the B-field part, is 
\be \la{vecSi}  
I = 2 \pi  \sqrt{ (w. w) (\tilde w . \tilde w) - ( w. \tilde w)^2  } + 2 \pi i B_{ij} w^i \tilde w^j 
\ee 

As an example, a worldsheet instanton that only wraps   the first torus gives
\be \la{SingIN}
I = 2 \pi (\rho_2 \pm i \rho_1 ) = 2\pi (Z \pm T)~,~~~~{\rm with} ~~~w^i = (1,0,0,0) ~,~~~~\tilde w^i = ( 0, \pm1,0,0) 
\ee 
This action is always positive within the dotted lines of figure \nref{T4WH}. 

 However, a more complicated worldsheet instanton can have a more interesting action. A simple example is given by the choices 
 \be \la{vecCho}
 w^i = (1,0,0,-1) ~,~~~~~ \tilde w^i = (0,1,1,0) 
 \ee 
 In this case, the action is 
 \be \la{ActIN}
 I = 2 \pi \left( 1+\rho_2^2 + \rho_1^2   + 2 i \rho_1 \right) = 4\pi  \left( 1- \tan( \tau/2) \right)
 \ee 
 where the last term comes from the $B$-field. We  also have set the values to lie along a geodesic given by \be \la{GeoCaK}
 Z=\rho_2 =   { 1 \over \cos (\tau/2) } ~,~~~~~~i T = \rho_1= i  \tan(\tau/2)
 \ee 
 This shows that the instanton becomes zero at $\tau = \pi/2$, which is a distance smaller than $\tau_{\text{IDB}} = {\pi \over 2} \sqrt{ 5/2}$. 

\begin{figure}[h]
	\begin{subfigure}{.5\textwidth}
		\centering
		\includegraphics[width = 7.0cm]{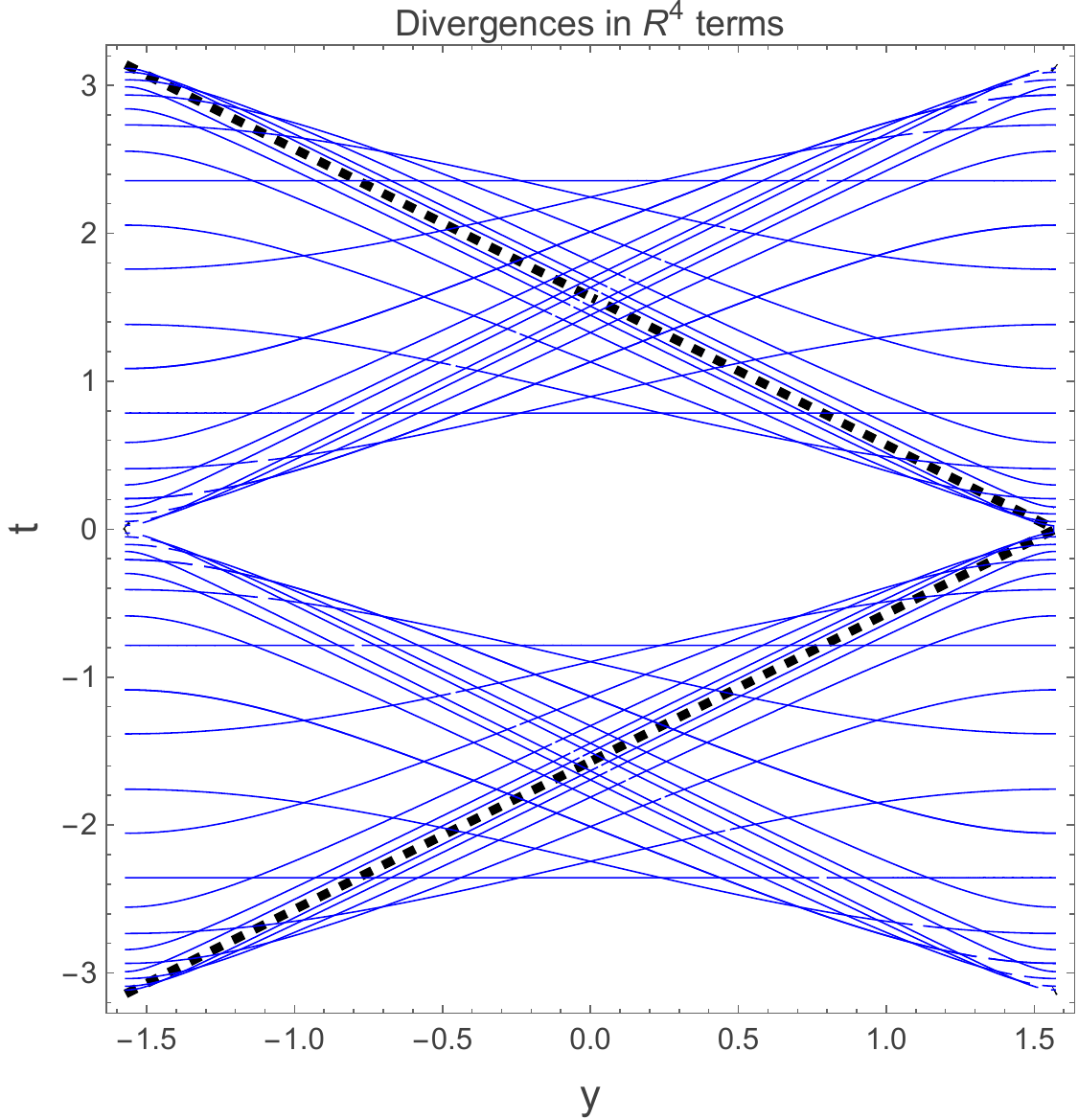}
	\end{subfigure}
    	\begin{subfigure}{.5\textwidth}
		\centering
		\includegraphics[width = 7.0cm]{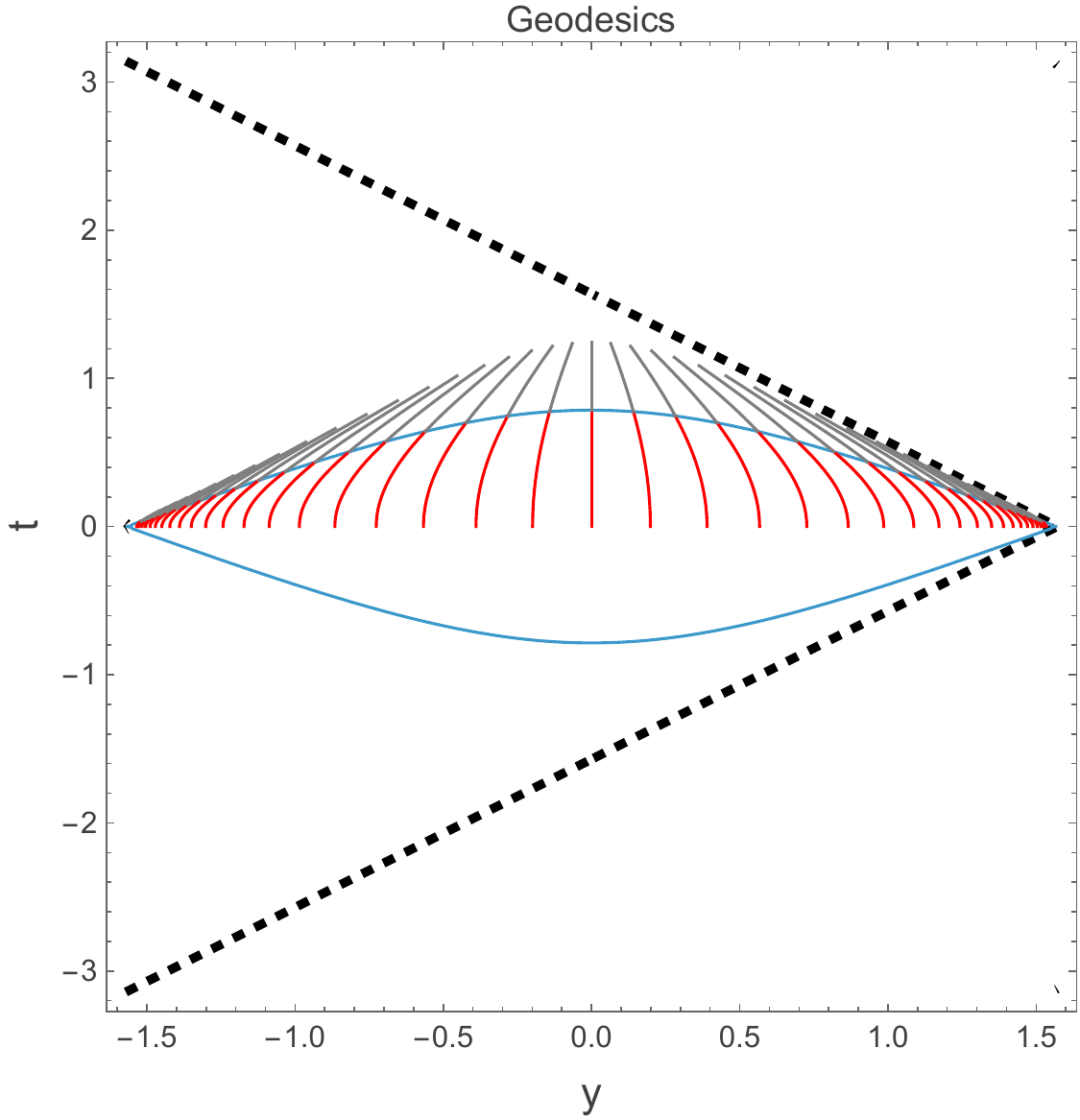}
	\end{subfigure}
   \begin{center} (a) ~~~~~~~~~~~~~~~~~~~~~~~~~~~~~~~~~~~~~~~~~~~~~~~~~~~~(b)  \end{center}
    \caption{Moduli space in the global coordinates from \nref{eq:global}. (a) blue lines are locations where $R^4$ corrections diverge due to an actionless instanton / massless state.  We plotted $I=0 $ in \nref{ActInK} with $r \in \{ 1/32, 1/16, 1/8, ..., 16, 32 \}$. The dotted lines are where the simplest instanton \nref{SingIN} becomes zero.  (b) time-symmetric geodesics with differing values of $\gamma$. The red portion is $\tau \in (0, \pi / 2)$ and the gray portion is the rest of the geodesic. We see that they always cross the massless envelope at a distance of    $\pi/2$  \nref{TauStar}. }
    \label{fig:T4geos}
\end{figure}

  More generally, we can consider the worldsheet instantons given by 
 \be \la{vecChoGe}
 w^i = p (r,0,0,-1) ~,~~~~~ \tilde w^i =  q (0, r^{-1} ,1,0) ~,~~~~~
 \ee 
 with rational $r$ and appropriate integers $p$ and $q$.
This leads to the action
\be \la{ActInK}
I = 2 \pi |pq| \left[ r + { Z^2 -T^2 \over r } 
\pm 2 T \right] ~,~~~~~\rho_2 = Z ~,~~~~~\rho_1 = i T 
\ee 
where the $\pm$ is determined based on whether the relative sign of $p$ and $q$ is negative or positive. Setting this to zero for various rational values of $r$ we get the blue curves in figure \nref{fig:T4geos}(a).
These are the curves where the action of some instanton becomes zero. The first place where there is {\it some} instanton with zero action forms the envelope curve in blue in figure \nref{fig:T4geos}, given by  
\be \la{Envel}
 Z = \sqrt{2}|T| 
 \ee 
 This is found from \nref{ActInK} by extremizing over $r$ and setting the resulting action to zero. 
 This curve lies at a distance of $\pi/2$ from the $T=0$ line, as can be checked as follows. First, this is true for a particular point reached by the geodesic \nref{GeoCaK}. Then the scaling symmetry of AdS$_2$ in Poincaré coordinates, and the equation \nref{Envel}, ensures it is true for the rest of the points. 

The corrections due to these instantons become large. In particular, we can discuss their impact on the higher derivative corrections. 

\paragraph{$R^4$ corrections} 
As we did for the IIB wormholes before, we can also see that there is an issue with the $R^4$ higher derivative corrections to supergravity, which were studied  
 in \cite{Obers:1999um} for toroidal compactifications.
The coefficient of the $R^4$ term is conjectured to be the Eisenstein series for the full U-duality group, generalizing the case of~\eqref{eq:SL2Eisenstein} involving the $SL(2, \Z)$ duality group. This coefficient takes the form
\begin{align}
    f_{R^4} = 2 \zeta(3) \frac{V_d}{g_s^2} + I_d + \text{non-perturbative}
\end{align}
The tree-level part is obtained by dimensionally reducing the 10d tree-level part -- this does not blow up for geodesics which never reach the weak-coupling horizon. So we can focus on the one-loop part, 
\begin{align}
    I_d = 2 \pi \int \frac{d^2 \tau}{\tau_2^2} Z_{d,d}(G, B, \tau) \, ,
\end{align}
where $\tau$ is the modular parameter of the torus worldsheet, and $Z$ is the partition function corresponding to the even self-dual lattice describing the compactification on $T^d$, given by
\begin{align} \la{MomWind}
    Z_{d,d} \ = \ V_d \sum_{w^i, \tilde{w}^i} e^{-\frac{\pi}{\tau_2} (w^i + \tau \tilde{w}^i)(G_{ij} + B_{ij})(w^i + \bar{\tau} \tilde{w}^i)} = (\tau_2)^{d/2} \sum_{n_i, w^i} e^{-\pi \tau_2 M^2 - 2 \pi i \tau_1 n_i w^i}
\end{align}
 where $V_d = \sqrt{G}$ is the volume of the torus and $M^2$ is defined in \nref{MassSqu}.
These two expressions are related by Poisson resummation. On the left, we have $w^i$ and $\tilde w^i$ represent windings of the torus worldsheet around the internal cycles. After Poisson resummation, one of these becomes a momentum $n_i$. The exponent in the left expression is proportional to the action of the wound string; in the right expression we see that $\tau_2$ multiplies mass squared while the $\tau_1$ integral multiplies the level matching condition, forcing $n_i w^i$ to zero. 

If we use the all-winding representation of \nref{MomWind}, then we will ultimately find that the problem is caused by the exact instantons we discussed above in \nref{vecChoGe}. This calculation is detailed in \ref{app:modulispaceintegral}. It might also be instructive to see the issue in the Poisson dual ``Hamiltonian'' frame on the right of \nref{MomWind}. We can do this by directly computing the mass for arbitrary winding and momentum vectors. If we find one where both the mass-squared and level-matching condition are zero at some location in moduli space $G, B$, then the higher-derivative corrections blow up when the scalars in the wormhole background reach that location. The mass matrix is given by
\begin{align}
   M^2 =  v_i G^{ij} v_j + w^i G_{ij} w^j \, ~~~~{\rm with } ~~ v_i \equiv  n_i + B_{ij} w^j \la{MassSqu}
\end{align}
Now we will take the ``diagonal path'' that was considered in \cite{Arkani-Hamed:2007cpn}, where all of the dilatons are equal and all of the axions equal. The $T^4$ compactification has 4 complex scalars. To trace out the same total distance in moduli space, each axi-dilaton system only has to move half as far. So we find that to make a wormhole, the imaginary distance traversed in each axi-dilaton $H_2$ space is 
\begin{align}
    \frac{\tau_\text{IDB}}{2} = \frac{\pi}{4} \sqrt{\frac{2(D-1)}{(D-2)}} = \frac{\pi}{4} \sqrt{\frac{5}{2}}
\end{align}
Now we will analyze geodesics on the effectively 2d moduli space which move this distance. Looking at the geodesics \nref{eq:scalareqgeneral} and choosing integration constants for a time symmetric geodesic we find 
\begin{align} \la{GeoExp}
    Z = e^{\gamma} { 1 \over \cos (  \tau / 2  ) } \, , \qquad T = e^{\gamma} \tan(\tau /2) \, .
\end{align}
Where the argument is now $\tau / 2$ because each field is tracing out half of the total distance $\tau$. Then we can show that in fact, the further distance that can be traveled is $\tau_\text{max} = \pi /4$ before hitting a massless state.
Take, for instance, the following choice: 
\begin{align}
    \label{eq:mystate}
    w = (w, 0,0, w k^{-1}) \, ,\ \qquad n = (-w,0,0,wk) \, .
\end{align}
Level matching is satisfied manifestly and if we pick $k = e^\gamma$ then we see that 
\begin{equation}
    M^2 = 4 w^2 \cos(\tau) \, ,
\end{equation}
which goes from positive to negative at $\pi / 2$. Thus we have some particular massless states, which depend on $\gamma$. Now, to do this, we assumed that $e^\gamma$ was rational. If it is not, then we will not find a massless state exactly at 
\be \la{TauStar}
\tau_* = \pi / 2 
\ee 
but we can make the massless state arbitrarily close by approximating $e^{\gamma}$ more and more closely by $w_2 / w_1$. We can check numerically that changing $e^\gamma \to (1 + \epsilon) e^\gamma$ in the above solution always has a small effect on the point where $M^2 = 0$, always moving it to infinitesimally larger $\tau$.    If we go beyond these massless points, then the mass squared will become negative -- the states become tachyonic. 

The authors of \cite{Arkani-Hamed:2007cpn} emphasized that their wormholes could then be used in AdS$_3 \times S^3 \times T^4 $ compactifications. Actually, to be precise, in order to construct this AdS$_3$ wormhole, we need to make sure that the fluxes we introduce keep the scalar fields massless. This is the case if we choose NS 3-form flux on $S^3$ and AdS$_3$. There are two points we can make now. First,  the wormhole now could be an AdS$_3 $ wormhole,  and the distance could be just 
$\tau = \pi/2$, which is the value of $\tau'_{IDB}$ for AdS$_3$. This is precisely the imaginary distance where the massless (soon to be tachyonic) states appear \nref{TauStar}.
Second, we can ask whether the boundary theory faces a similar problem at this point. The boundary theory is a deformation of the symmetric product $Sym(T^4)^{Q_1 Q_5} $. At precisely the symmetric point and for the same values of the $T^4$ moduli, we see that   $M^2$ in \nref{MassSqu} is setting the values of the dimensions of momentum and winding modes on the torus. Therefore the fact that they are crossing zero implies that the theory has states that violate the unitarity bound.  Of course, we discussed these states only at the symmetric orbifold point, and it would be nice to see that this problem remains after turning on the deformation away from the symmetric product point. 

The paper \cite{Arkani-Hamed:2007cpn} had another interesting wormhole example involving a heterotic or type II theory compactified on a six-dimensional Calabi Yau to four non-compact dimensions. In this case, the four-dimensional dilaton and $B$ field give rise to an axion dilaton pair, with an  $H_2$ moduli space. There is another  $H_2$ factor from the volume and NS $B$ field proportional to the Kahler form in the internal space. The first has radius one and the second radius $\sqrt{3} $ in our units \nref{ManFi}. This means that we can find a geodesic that is again within a Poincar\'e patch and which does not cross any obvious zero action instanton lines (like the red one in figure \ref{T4WH}). We also expect to find a problem with the compactification along this trajectory, but we have not managed to do so. Further wormhole solutions have been discussed in \cite{Hertog:2017owm, Katmadas:2018ksp, VanRiet:2020pcn, Loges:2023ypl}. It would be nice to understand if these also suffer from actionless instantons. We leave this to the future. 

More generally, in Calabi-Yau compactifications, we can have hundreds of moduli. With hundreds of axions, we can consider a wormhole where each axion moves only a small distance. The IDB conjecture \nref{NDWC} would then imply that there should be a problem even in this case. The fact that these wormholes would naively have no problem is similar to the arguments for inflation with $N$ fields in \cite{Dimopoulos:2005ac}, and this could offer an interesting test of the IDB conjecture.

\section{Discussion}

In this paper, we have reviewed wormholes constructed using imaginary scalar fields in flat space and in AdS, and used their existence to motivate an ``imaginary distance bound'' on the maximal distance one can move in an imaginary direction in scalar field moduli space. The main idea is that as we increase the distance in moduli space, eventually we reach a point where   wormholes become unsuppressed and give rise to large deviations in the effective field theory.

In some string theory examples, we found a specific mechanism in the theory defined within a single asymptotic region that explains why a problem arises as we move in the imaginary direction. This sometimes occurs before the imaginary distance bound. 

In type IIB string theory, where wormholes are formed using the string theory dilaton and the Ramond-Ramond zero form, the wormhole trajectories must cross a region in moduli space where $D(-1)$-instanton action becomes zero, and hence unsuppressed in the bulk. The $R^4$ corrections diverge at this point, giving large corrections to the equations of motion. For ten-dimensional string theory compactified on $T^4$, we found that the wormhole solutions have regions where worldsheet instantons become actionless, also giving a divergent one-loop contribution to the $R^4$ correction.
Besides giving evidence for our conjecture, these effects mean that the wormhole solution of low energy gravity is no longer reliable. The reason is that, in the wormhole solution, the asymptotic values of the scalars are imaginary at precisely the imaginary distance bound.  Therefore, the IDB \nref{NDWC} prevents trustworthy flat space scalar field wormholes, in a way that is analogous to how the weak gravity conjecture prevents stable extremal black holes. In fact, we saw in section \ref{DimReCh} that the latter is a special case of the former.

We also discussed several wormhole solutions that can be constructed from dimensional reductions. By reducing pure gravity, one can form wormholes by allowing the KK scalar to be imaginary. In fact, these wormholes have the same metric as the Schwarzschild solution in the higher-dimensional theory and differ only by the choice of contour in the complex $r$-plane. They are essentially the same as the double cone wormholes discussed in \cite{Saad:2018bqo}. A similar result is obtained by dimensionally reducing the Einstein-Maxwell theory. In that case, we found that the superextremal (self-repulsive) states predicted by the weak gravity conjecture become actionless instantons in the lower-dimensional theory, providing a further example where the IDB conjecture is satisfied.

Compactification can turn metric components into scalars in the lower-dimensional theory. This allowed us to consider the relation between our bound on imaginary scalar fields and the KSW condition \cite{Kontsevich:2021dmb, Witten:2021nzp} on complex metrics. For a general torus compactification we showed that the imaginary distance bound precisely agrees with the KSW condition applied to the \textit{internal metric components}, meaning that the wormholes are on the edge of being 
forbidden.
These examples show that the IDB is a generalization of the KSW criterion, at least with respect to compactification moduli. It also suggests that there may be a generalization of KSW that involves all fields -- not just the metric -- that is valid in more generality, including fields which are not moduli. 

We have argued for the Imaginary Distance Bound, in the context of gravity or holographic CFTs. It would be interesting to know whether similar bounds apply for more general CFTs, or even QFTs. QCD at imaginary $\theta$ has been studied, focusing on small $|\theta|$, in order to avoid the sign problem in numerical studies, see \textit{e.g.}  \cite{Vadacchino:2024ulq}.  

There are many interesting wormholes which are {\it not} scalar field wormholes, such as the wormholes that we get by adding matter or spatial inhomogeneities to the AdS boundary, see \textit{e.g.} \cite{Marolf:2021kjc,Maloney:2025tnn}. Another simple class of such wormholes involves empty Euclidean AdS space and hyperbolic space boundaries \cite{Maldacena:2004rf}, though  \cite{Witten:1999xp} argued that they do not arise in the common examples of AdS/CFT. The IDB \nref{NDWC}  is in the same spirit, but in the context of simple scalar field wormholes. 

It would also be nice to make a cleaner connection to the Swampland distance conjecture, which involves {\it real} distances, and leads to a bound if we impose a bound on the lightest massive spin two particle. Real distance bounds 
are particularly interesting during inflation, where they translate to a bound on the tensor-to-scalar ratio \cite{Lyth:1996im}, a quantity that will become increasingly constrained by observations in the near future \cite{Chang:2022tzj}. 

{\bf Acknowledgments }
We thank G. Di Ubaldo, L. Iliesiu, H. W. Lin, and C.   Yan for discussions on their very similar paper \cite{OtherPaper}. We also thank N. Arkani-Hamed, S. Caron-Huot, X. Dong,  C. Johnson, D. Harlow, S. Komatsu, J. Kudler-Flam,  H. Liu, D. Marolf,  B. Pioline, G. Shiu, A. Strominger, C. Vafa, C. Virally, E. Witten,  and Y. Zhao for helpful conversations.

The work of JM was supported in part by U.S. Department of Energy grant DE-SC0009988, and the Leinweber Forum at the IAS. 
The work of AM is supported in part by the Simons Foundation Grant No. 12574.  The work of BM is supported by the Gloria and Joshua Goldberg fellowship.

\appendix

\addtocontents{toc}{\protect\setcounter{tocdepth}{1}}

\section{Axions and the axion weak gravity conjecture}
\label{app:axions}

Our discussion is independent of any possible periodicity of the fields for real fields. 
In the case of periodic fields, or axions,  it is customary to write the action in terms of a periodic field 
$\tilde a  $ whose action is normalized as 
\be 
S =   f_a^{D-2 } \int \half (\nabla \tilde a )^2  ~,~~~~~~~\tilde a = \tilde a  + 2\pi 
\ee 
In terms of our normalizations in \nref{ManFi} we have 
\be 
\varphi = \tilde a  \sqrt{ 8 \pi G_N f_a^{D-2} }  
\ee 
and the imaginary distance bound \nref{IDB} is 
\be \la{IDBa}
{\Delta \tilde a  \over 2 }  = i {\pi \over 2 } \sqrt{ 2 (D-1)\over (D-2)} { 1 \over \sqrt{ 8 \pi G_N f_a^{D-2} }}
\ee 
which looks more complicated,  and it obscures the fact that the distance bound is completely independent of $f_a$ or the period of the axion. It physically depends only on the Planck mass.  Note that the wormhole action with fixed charge $n$ is $I_n = n \Delta \tilde a $. 

\subsection{The axion weak gravity conjecture}
\label{app:axionWGC}

The weak gravity conjecture has been applied to axions \cite{Rudelius:2015xta, Brown:2015iha, Montero:2015ofa}, with the conjecture being that there must be a ``superextremal instanton'' which would satisfy
\begin{align}
    \frac{n}{f^{(D-2)/2}} \gtrsim S_{\text{inst}} \kappa \, ,
\end{align}
where $2 \kappa^2 = 16 \pi G_N$. 
The formulation \nref{NDWC} of the Imaginary Distance Bound suggests that there should be some charge $n$ and an instanton whose action is smaller or equal to half the wormhole action (for fixed charge). This translates into 
\be 
 { S_{\text{inst}, n } \over n } \leq {  I_n \over 2 n } =   {\pi \over 2} \sqrt{ 2 (D-1)\over (D-2)} { 1 \over \sqrt{ 8 \pi G_N f_a^{D-2} }}
\ee 
Here,  $S_\text{inst}$ does not include the coupling between the charge and axion. This bound is discussed at more length in \cite{OtherPaper}. This form of the bound was proposed in \cite{Montero:2015ofa} based on spacetime wormholes.

\section{The KSW criterion for the Schwarzschild double cone }
\la{KSWDC}

The $D+1$-dimensional Schwarzschild geometry
\be
ds^2 = -f(r) dt^2 + \frac{dr^2}{f(r)} + r^2 d\Omega_{D-1}^2
\ee
can be easily complexified by deforming the radial coordinate $r$ from the real axis to some contour in the complex $r$ plane.  Letting $r=r(u)$ for a real coordinate $u$, the (complex) metric is then
\be\label{complexSch}
ds^2 = -f(r(u)) dt^2 + \frac{r'(u)^2}{f(r)}du^2 + r(u)^2 d\Omega_{D-1}^2
\ee
We are interested in SSS-type contours, such as the red contour in Figure \ref{PlotContour}, which avoid the zero of $f(r)$.  We want to know whether the KSW criterion
\be\label{kswSch}
|Arg(f(r))| + |Arg(r'(u)^2)/f(r(u))| + (D-1) |Arg(r^2)| \le \pi
\ee
for the metric \nref{complexSch} is satisfied for such a contour.
This answer is no, as was noted already in \cite{Chakravarty:2024bna}; we provide here a streamlined argument.

To begin, note that there must be a point along the red contour in Figure \ref{PlotContour} where $r'(u)$ is purely real; this occurs when $\text{Im}(r(u))$ is extremized so $\text{Im}(r'(u)) = 0$.  At this point $r'(u)$ drops out of the KSW criterion.  We then note that for any non-zero, complex $z=f(r(u))$
\be
|Arg(-z)| + |Arg(1/z)| = \pi
\ee
Equation \nref{kswSch}  then implies that $r(u)$ is real.  Our conclusion is that, once $r$ is given a non-zero imaginary part, this imaginary part must increase or decrease monotonically.  As a result the KSW criterion will never be satisfied for a contour that asymptotes to the real $r$ axis at both ends.

\section{Wormhole in JT gravity }
\la{WHJT}

Here we consider a JT gravity action plus a scalar field 
\be 
I = I_{top} - { 1\over 4 \pi } \int \sqrt{g} \phi (R +2 ) + \int \sqrt{g} \half (\nabla \varphi )^2  ~,~~~~~~~I_{top} = { S_0 \over 4 \pi } \int \sqrt{g} R 
\ee 
The equations of motion for $\phi$  imply that we have a Euclidean AdS$_2$ spacetime 
\be \la{MetCosGl}
ds^2 = { { d t^2 + d{\tilde \tau }^2 \over \cos^2 \tilde \tau} }
\ee 
The equations of motion for the scalar field in this metric are the same as on a flat strip,  and there is a simple linear solution of the form 
\be \la{LinFie}
\varphi(\tilde \tau )  =   { 2  \over \pi } \tilde \tau \varphi_b  ~~~~~~~~~ \varphi ( \pm {\pi \over 2} ) = \pm \varphi_b
\ee 
where we take $\varphi_b$ to be imaginary. 
The equations of motion for the metric imply a series of equations for the dilaton which can be solved by \cite{Garcia-Garcia:2020ttf}\footnote{The profile for the dilaton is also the same as the solution found in appendix C.3 of \cite{Maldacena:2018lmt}, where the negative energy was coming from a Casimir energy. It is the same because the negative energy in both cases has the same form, it is constant in the strip coordinates.}
\be \la{DilAdS2}
\phi = - { 4 \varphi_b^2 \over \pi } \left( 1 + \tilde \tau \tan \tilde \tau \right)  
\ee 
Note that the dilaton grows towards the boundary only if $\varphi_b$ is imaginary.

The renormalized action is $I =  {\varphi_b^2 \over \pi } \beta $, where $\beta$ is the total length of the Euclidean time direction. Note that it is negative for imaginary $\varphi_b$.

\section{More details on the wormhole related to charged black holes }
\la{AppCharged}

Here we give more details on the solution discussed in section \nref{ChRegWH}. 
We start with the action \nref{MetMOs}. We consider a standard wormhole which involves a geodesic in the moduli space. This geodesic is parametrized by the proper time $\tau$. The geodesic is given by \nref{GeoReCh} or 
 \be 
 e^{  \nu } = {    \cos( {\pi \over 2} \cos \epsilon ) \over \cos ( \tilde \tau \cos \epsilon )} ~,~~~~ \tilde \alpha  = i  \cos( {\pi \over 2} \cos \epsilon ) \tan(  \tilde \tau \cos \epsilon ) ~,~~~~~\varphi= i \tilde \tau \sin \epsilon  ~,~~~~~\tau = \hat \kappa \tilde \tau 
 \ee 
 where  $- \pi/2 \leq \tilde \tau \leq \pi/2 $. We have adjusted the value of $\gamma$ in \nref{GeoReCh} so that $\nu =0 $ at the wormhole endpoints. 
 The solution for $a$ and $\tilde \tau$ is the same as the flat space wormhole in \nref{MeWH} \nref{FaWH} except that we write it in terms of conformal coordinates, as in \nref{ConfWH}, which we repeat here for convenience 
 \be 
 ds^2_{D}  = a^2 \left( { d \rho^2 \over (D-2)^2 } + d\Omega_{D-1}^2 \right) ~,~~~~~~ { a^{D-2} \over a_0^{D-2} } = \cosh\rho = { 1 \over \cos \tilde \tau } 
 \ee 
 We now uplift this metric to $D+1$ dimensions, and we express the solution in terms of the variable $\tilde \tau $, using ${ d\rho \over d\tilde \tau } =   { 1 \over \cos \tilde \tau } $, 
 \bea \la{MEtCH}
 ds^2 &=& e^{ 2 \nu } dt^2 +  (e^{-2 \nu})^{ 1 \over D-2} a^2 \left( { d \rho^2 \over (D-2)^2 } + d\Omega_{D-1}^2 \right)
 ~,~~~~~~ A_t \propto \tilde \alpha  ~,~~~~~\varphi = i \tilde \tau \sin \epsilon  
 \\ \la{MetHDi} ds^2 &=& \left[ { \cos( {\pi \over 2 } \cos \epsilon ) \over \cos\left(\tilde \tau \cos \epsilon    )  \right)}  \right]^2 dt^2 + 
 a_0^2 \left[ {   \cos\left( \tilde \tau \cos \epsilon    \right) \over  \cos \tilde \tau \cos ( {\pi \over 2 } \cos \epsilon )   } \right]^{ 2 \over D-2}    \left( { d {\tilde \tau }^2 \over (D-2)^2 \cos^2 \tilde \tau } + d\Omega_{D-1}^2 \right) ~~~~~~~~~ 
 \eea 
 Note that $a_0$ is an overall rescaling of the $(D+1)$-dimensional solution. For small $\epsilon$ we will show that this solution has a central region  where it takes a nearly AdS$_2$ form as the wormhole in appendix \ref{WHJT}.
 
 To get oriented, it is useful to write the values of    
  $g_{tt}$ and the radius of the sphere at $\tilde \tau =0$  
\be 
g_{tt} =\cos^2( {\pi \over 2 } \cos \epsilon)  \sim { \pi^2 \epsilon^4 \over 16 } ~,~~~~~~~~~~~~r_{e}^{D-2}  = { a_0^{D-2}  \over   \cos  ( {\pi \over 2 } \cos \epsilon)    } \sim { 4 a_0^{D-2}   \over \pi^2 \epsilon^2 }
\ee
Clearly, we need to take $a_0$ very small in order to have an extremal radius $r_e$  of order one.

 The transition between the near horizon region and the asymptotically flat region occurs at a value of $\tilde \tau  $   near $\pi/2$ where 
 \be 
\cos \tilde \tau \sim {  \epsilon^2 } ~,~~~~~~{\rm or } ~~~~~ ({ \pi \over 2} -\tilde \tau ) \sim \epsilon^2 
\ee 
This is obtained by looking at 
\be 
 \tilde \tau \cos \epsilon        \sim    { \pi \over 2 }  - { \pi \epsilon^2 \over 4 }  -  \left( {\pi \over 2 } -\tilde \tau \right)  
\ee
and demanding that the last two terms are comparable. Inserting this in the expression for $g_{tt} $ we find 
\be 
g_{tt} = \left[ { {   \pi \epsilon^2 \over 4}  \over  { \pi \epsilon^2 \over 4} +  \cos \tilde \tau } \right]^2 ~.~~~~~~~~~
\ee 
We see that when $\cos \tilde \tau \ll \epsilon^2 $ it is approximately one, and the second term in \nref{MetHDi} becomes the flat space metric.  (We need to use that $( { \pi \over 2 } - \tilde \tau ) \propto  1/r^{D-2} $ where $r$ is the usual flat space radial coordinate). 
On the other hand, when $ \cos \tilde \tau \gg \epsilon^2 $ we can approximate the metric by 
\be 
ds^2_{D+1} = r_e^2 \left[ { 1 \over (D-2)^2 } \left({ d\hat t^2  + d\tilde \tau^2 \over \cos^2 \tau }\right) + d\Omega_{D-1}^2 \right]
\ee 
We see that the metric \nref{MEtCH} takes the $AdS_2 \times S^{D-1}$ form. This is similar to what we found around \nref{BRSol}. 

We can also find something that we can interpret as the dilaton of JT gravity by expanding the radius of the sphere to the next order in $\epsilon^2$ in the region $\cos \tilde \tau \gg \epsilon^2$. We find  
\be 
r_e^2 \left(1 + { \epsilon^2  \over (D-2) } \tilde \tau \tan \tilde \tau  + \cdots \right)
\ee 
where the $\epsilon^2 $ term  has the same form as the dilaton in \nref{DilAdS2} (up to an additive constant). Note also that  $\varphi $ in \nref{MEtCH}  is also linear in $\tilde \tau $  as in \nref{LinFie}.

\section{$T^4$ moduli space integral}
\label{app:modulispaceintegral}

In section \ref{sec:UV}, we showed that the $R^4$ corrections of type II string theory compactified on $T^4$ blow up due to the presence of states that become massless on the wormhole background. Here we will perform the same calculation in the Poisson dual frame, where the corrections are due to instantons rather than massless particles. The goal here will be to evaluate
\begin{align} \la{StaAp}
    I_4 = 2 \pi V_4 \int_{\mathcal F} \frac{d^2 \tau}{\tau_2^2}  \sum_{w^i, \tilde{w}^i} e^{-\frac{\pi}{\tau_2} (w^i + \tau \tilde{w}^i)G_{ij} (w^i + \bar{\tau} \tilde{w}^i) +2 \pi i   w^i \tilde w^j B_{ij})}
\end{align}
Now we need to unfold this integral, which means to turn the integral over $\mathcal F$ into an integral on the entire upper half plane. We do this by separating the $4 \times 2$ matrix $(w, \tilde w)$ into rank 0, rank 1, and rank 2 parts. 

For the rank $0$ part, the windings are all zero and 
\begin{align}
    I_4^{(0)} = 2 \pi V_4 V_{\mathcal F}
\end{align}
where $V_{\mathcal F} = \pi / 3$ is the area of the fundamental domain. This is finite everywhere.

For the rank 1 part, we can define $w^i = c m^i$ and $\tilde w^i = \tilde c m^i$, and then sum over vectors $m^i$ and integers $c, \tilde c$. As a result the $B$ part drops out and we get
\begin{align}
    I_4^{(1)} = 2 \pi V_4 \sum_{m \neq 0} \int_{\mathcal F} \frac{d^2 \tau}{\tau_2^2} \sum_{(c, \tilde c) = 1} e^{-\frac{\pi}{\tau_2}|c + \tilde c \tau|^2 m^i G_{ij}m^j}
\end{align}
The integral of the sum over primitive integer pairs can be rewritten as the unfolded integral on the strip $\mathcal S$ and we get
\begin{align}
    I_4^{(1)} = 2 \pi V_4 \sum_{m \neq 0} \int_{\mathcal S} \frac{d^2 \tau}{\tau_2^2}  e^{-\frac{\pi}{\tau_2} m^i G_{ij}m^j} = 2 V_4 \sum_{m \neq 0} \frac{1}{m^i G_{ij} m^j}
\end{align}
This doesn't blow up for any particular choice of $m$ -- the real part of $G$ is symmetric so the real part of the denominator is manifestly positive. It has a divergence that is universal -- it is not the moduli space dependent divergence we are looking for -- and we think it should just be removed by a suitable subtraction. One can see that, in the Hamiltonian representation of \nref{MomWind}, this divergence arises from states with zero momentum and winding numbers, and it is due to the infrared part of the one-loop diagram \cite{Obers:1999um}. 

Now let's consider the rank 2 part. First we can write $W = (w, \tilde w)$ and use the identity
\begin{align}
    \sum_{\text{rank(W) = 2}} = \sum_{W \in \mathcal R_2} \sum_{\gamma \in SL(2, \Z)} \, ,
\end{align}
where $\gamma$ acts on $W$ by right multiplication
\begin{align}
    \gamma = \begin{pmatrix}
        a & b \\ c &d
    \end{pmatrix} \, , \qquad W \cdot \gamma = (a w - b \tilde w, -c w + d \tilde w)
\end{align} and $\mathcal{R}_2$ are any set of representatives of rank two matrices under these orbits. As a result we can use the sum over $\gamma$ to unfold the integral, giving
\begin{align}
    I^{(2)}_4 = 2 \pi V_4 \sum_{W \in \mathcal R_2} \int_{\mathbb H} \frac{d^2 \tau}{\tau_2^2}   e^{-\frac{\pi}{\tau_2} (w^i + \tau \tilde{w}^i)G_{ij} (w^i + \bar{\tau} \tilde{w}^i) +2 \pi i   w^i \tilde w^j B_{ij}}
\end{align}
Defining $a \cdot b = a_i G^{ij} b_i$, the exponent equals 
\begin{align}
    -\frac{\pi}{\tau_2}  \left( 
    \tau_1^2 \tilde w \cdot \tilde w + 2 \tau_1 w \cdot \tilde w + w \cdot w + \tau_2^2 \tilde w \cdot \tilde w \right) + 2 \pi i   w^i \tilde w^j B_{ij}
\end{align}
so the $\tau_1$ integral gives
\begin{align}
    I^{(2)}_4 = 2 \pi V_4 \sum_{W \in \mathcal R_2} \int_0^\infty \frac{d \tau_2}{\tau_2^2}   e^{-\frac{\pi}{\tau_2 \tilde w \cdot \tilde w} (\tau_2^2 (\tilde w \cdot \tilde w)^2  + (\tilde w \cdot \tilde w) (w \cdot w)- (w \cdot \tilde w)^2)} e^{2 \pi i   w B \tilde w } \sqrt{\frac{\tau_2}{\tilde w \cdot \tilde w}}
\end{align}
Then we can do the integral over $\tau_2$ to get
\begin{align}
    I^{(2)}_4 = 2 \pi V_4 \sum_{W \in \mathcal R_2}  \frac{e^{-2 \pi \sqrt{(\tilde w \cdot \tilde w)(w \cdot w) - (w \cdot \tilde w)^2} }}{\sqrt{(\tilde w \cdot \tilde w)(w \cdot w) - (w \cdot \tilde w)^2}} e^{2 \pi i   w B \tilde w } \la{SumWindI}
\end{align}
The denominator turns out to never hit zero for rank two states. So there is not a single state that is responsible for the blow-up. Still it can be useful to look at some particular choice of states. One interesting one is $w = p(1,0,0,-1)$, $\tilde w = q(0,1,1,0)$. Then the summand becomes
\begin{align} \la{InstWind}
    \frac{e^{-2 \pi |p q|(1 + \rho_1^2 + \rho_2^2 + 2 i \rho_1)}}{|p q|  (1 + \rho_1^2 + \rho_2^2)} \ = \ \frac{e^{-4 \pi |p q|(1 - \tan \sigma)}}{ 2 |p q| }
\end{align}
From this we can see the problem $\sigma = \pi / 4$. The exponent vanishes and the sum over $p$ and $q$ is divergent. This is the moduli dependent blow-up we are looking for and it comes from the same worldsheet instantons that we discussed in section \ref{sec:UV}.

\section*{}
\bibliography{cite.bib}

@article{Witten:2026twr,
    author = "Witten, Edward",
    title = "{Duality and Axion Wormholes}",
    eprint = "2601.01587",
    archivePrefix = "arXiv",
    primaryClass = "hep-th",
    month = "1",
    year = "2026"
}

@article{Chakravarty:2024bna,
    author = "Chakravarty, Joydeep and Maloney, Alexander and Namjou, Keivan and Ross, Simon F.",
    title = "{A new observable for holographic cosmology}",
    eprint = "2407.04781",
    archivePrefix = "arXiv",
    primaryClass = "hep-th",
    doi = "10.1007/JHEP10(2024)184",
    journal = "JHEP",
    volume = "10",
    pages = "184",
    year = "2024"
}

@article{Ivo:2026ijv,
    author = "Ivo, Victor and Tang, Haifeng",
    title = "{One-loop aspects of de Sitter axion wormholes}",
    eprint = "2603.02335",
    archivePrefix = "arXiv",
    primaryClass = "hep-th",
    month = "3",
    year = "2026"
}

@article{Chang:2022tzj,
    author = "Chang, Clarence L. and others",
    title = "{Snowmass2021 Cosmic Frontier: Cosmic Microwave Background Measurements White Paper}",
    eprint = "2203.07638",
    archivePrefix = "arXiv",
    primaryClass = "astro-ph.CO",
    reportNumber = "FERMILAB-PUB-22-309-PPD",
    month = "3",
    year = "2022"
}

@article{Lyth:1996im,
    author = "Lyth, David H.",
    title = "{What would we learn by detecting a gravitational wave signal in the cosmic microwave background anisotropy?}",
    eprint = "hep-ph/9606387",
    archivePrefix = "arXiv",
    reportNumber = "LANCASTER-TH-9612",
    doi = "10.1103/PhysRevLett.78.1861",
    journal = "Phys. Rev. Lett.",
    volume = "78",
    pages = "1861--1863",
    year = "1997"
}

@article{Dimopoulos:2005ac,
    author = "Dimopoulos, S. and Kachru, S. and McGreevy, J. and Wacker, Jay G.",
    title = "{N-flation}",
    eprint = "hep-th/0507205",
    archivePrefix = "arXiv",
    reportNumber = "SLAC-PUB-11016, SU-ITP-05-08",
    doi = "10.1088/1475-7516/2008/08/003",
    journal = "JCAP",
    volume = "08",
    pages = "003",
    year = "2008"
}

@article{Saad:2021rcu,
    author = "Saad, Phil and Shenker, Stephen H. and Stanford, Douglas and Yao, Shunyu",
    title = "{Wormholes without averaging}",
    eprint = "2103.16754",
    archivePrefix = "arXiv",
    primaryClass = "hep-th",
    doi = "10.1007/JHEP09(2024)133",
    journal = "JHEP",
    volume = "09",
    pages = "133",
    year = "2024"
}

@article{Mertens:2022irh,
    author = "Mertens, Thomas G. and Turiaci, Gustavo J.",
    title = "{Solvable models of quantum black holes: a review on Jackiw{\textendash}Teitelboim gravity}",
    eprint = "2210.10846",
    archivePrefix = "arXiv",
    primaryClass = "hep-th",
    doi = "10.1007/s41114-023-00046-1",
    journal = "Living Rev. Rel.",
    volume = "26",
    number = "1",
    pages = "4",
    year = "2023"
}

@article{Vadacchino:2024ulq,
    author = "Vadacchino, Davide and Bonanno, Claudio and Bonati, Claudio and Papace, Mario",
    title = "{The imaginary-$\theta$ dependence of the SU($N$) spectrum}",
    eprint = "2411.14022",
    archivePrefix = "arXiv",
    primaryClass = "hep-lat",
    doi = "10.22323/1.466.0399",
    journal = "PoS",
    volume = "LATTICE2024",
    pages = "399",
    year = "2025"
}

@article{Maloney:2025tnn,
    author = "Maloney, Alexander and Meruliya, Viraj and Van Raamsdonk, Mark",
    title = "{Ordinary wormholes}",
    eprint = "2503.12227",
    archivePrefix = "arXiv",
    primaryClass = "hep-th",
    month = "3",
    year = "2025"
}

@article{Marolf:2021kjc,
    author = "Marolf, Donald and Santos, Jorge E.",
    title = "{AdS Euclidean wormholes}",
    eprint = "2101.08875",
    archivePrefix = "arXiv",
    primaryClass = "hep-th",
    doi = "10.1088/1361-6382/ac2cb7",
    journal = "Class. Quant. Grav.",
    volume = "38",
    number = "22",
    pages = "224002",
    year = "2021"
}

@article{Witten:1999xp,
    author = "Witten, Edward and Yau, Shing-Tung",
    editor = "D'Hoker, Erik and Phong, Duong and Yau, Shing-Tung",
    title = "{Connectedness of the boundary in the AdS / CFT correspondence}",
    eprint = "hep-th/9910245",
    archivePrefix = "arXiv",
    doi = "10.4310/ATMP.1999.v3.n6.a1",
    journal = "Adv. Theor. Math. Phys.",
    volume = "3",
    pages = "1635--1655",
    year = "1999"
}

@article{Grimm:2018ohb,
    author = "Grimm, Thomas W. and Palti, Eran and Valenzuela, Irene",
    title = "{Infinite Distances in Field Space and Massless Towers of States}",
    eprint = "1802.08264",
    archivePrefix = "arXiv",
    primaryClass = "hep-th",
    doi = "10.1007/JHEP08(2018)143",
    journal = "JHEP",
    volume = "08",
    pages = "143",
    year = "2018"
}

@article{Garcia-Garcia:2020ttf,
    author = "Garc{\'\i}a-Garc{\'\i}a, Antonio M. and Godet, Victor",
    title = "{Euclidean wormhole in the Sachdev-Ye-Kitaev model}",
    eprint = "2010.11633",
    archivePrefix = "arXiv",
    primaryClass = "hep-th",
    doi = "10.1103/PhysRevD.103.046014",
    journal = "Phys. Rev. D",
    volume = "103",
    number = "4",
    pages = "046014",
    year = "2021"
}

@article{Held:2026huj,
    author = "Held, Jesse and Kaplan, Molly and Marolf, Donald and Wang, Zhencheng",
    title = "{Axion Wormholes and the AdS/CFT Factorization Problem}",
    eprint = "2601.02507",
    archivePrefix = "arXiv",
    primaryClass = "hep-th",
    month = "1",
    year = "2026"
}

@article{Held:2026bbo,
    author = "Held, Jesse and Kaplan, Molly and Marolf, Donald and Wang, Zhencheng",
    title = "{Lorentzian Path Integrals and Jackiw-Teitelboim wormholes with imaginary scalars}",
    eprint = "2601.09932",
    archivePrefix = "arXiv",
    primaryClass = "hep-th",
    month = "1",
    year = "2026"
}

@article{BenettiGenolini:2026raa,
    author = "Benetti Genolini, Pietro and Janssen, Oliver and Murthy, Sameer",
    title = "{Allowable complex metrics and the gravitational index of AdS$_5$ black holes}",
    eprint = "2601.23197",
    archivePrefix = "arXiv",
    primaryClass = "hep-th",
    month = "1",
    year = "2026"
}

@article{Heidenreich:2015nta,
    author = "Heidenreich, Ben and Reece, Matthew and Rudelius, Tom",
    title = "{Sharpening the Weak Gravity Conjecture with Dimensional Reduction}",
    eprint = "1509.06374",
    archivePrefix = "arXiv",
    primaryClass = "hep-th",
    doi = "10.1007/JHEP02(2016)140",
    journal = "JHEP",
    volume = "02",
    pages = "140",
    year = "2016"
}

@article{Ooguri:2006in,
    author = "Ooguri, Hirosi and Vafa, Cumrun",
    title = "{On the Geometry of the String Landscape and the Swampland}",
    eprint = "hep-th/0605264",
    archivePrefix = "arXiv",
    reportNumber = "CALT-68-2600, HUTP-06-A017",
    doi = "10.1016/j.nuclphysb.2006.10.033",
    journal = "Nucl. Phys. B",
    volume = "766",
    pages = "21--33",
    year = "2007"
}

@article{Saad:2019lba,
      author         = "Saad, Phil and Shenker, Stephen H. and Stanford, Douglas",
      title          = "{JT gravity as a matrix integral}",
      year           = "2019",
      eprint         = "1903.11115",
      archivePrefix  = "arXiv",
      primaryClass   = "hep-th",
      SLACcitation   = "%%CITATION = ARXIV:1903.11115;%%"
}

@article{Arkani-Hamed:2007cpn,
    author = "Arkani-Hamed, Nima and Orgera, Jacopo and Polchinski, Joseph",
    title = "{Euclidean wormholes in string theory}",
    eprint = "0705.2768",
    archivePrefix = "arXiv",
    primaryClass = "hep-th",
    doi = "10.1088/1126-6708/2007/12/018",
    journal = "JHEP",
    volume = "12",
    pages = "018",
    year = "2007"
}

@article{Bergshoeff:2005zf,
    author = "Bergshoeff, Eric and Collinucci, Andres and Ploegh, Andre and Vandoren, Stefan and Van Riet, Thomas",
    title = "{Non-extremal D-instantons and the AdS/CFT correspondence}",
    eprint = "hep-th/0510048",
    archivePrefix = "arXiv",
    reportNumber = "UG-05-07, SPIN-05-30, ITP-UU-05-44",
    doi = "10.1088/1126-6708/2006/01/061",
    journal = "JHEP",
    volume = "01",
    pages = "061",
    year = "2006"
}

@article{Loges:2023ypl,
    author = "Loges, Gregory J. and Shiu, Gary and Van Riet, Thomas",
    title = "{A 10d construction of Euclidean axion wormholes in flat and AdS space}",
    eprint = "2302.03688",
    archivePrefix = "arXiv",
    primaryClass = "hep-th",
    reportNumber = "KEK-TH-2495",
    doi = "10.1007/JHEP06(2023)079",
    journal = "JHEP",
    volume = "06",
    pages = "079",
    year = "2023"
}

@article{Hertog:2017owm,
    author = "Hertog, Thomas and Trigiante, Mario and Van Riet, Thomas",
    title = "{Axion Wormholes in AdS Compactifications}",
    eprint = "1702.04622",
    archivePrefix = "arXiv",
    primaryClass = "hep-th",
    doi = "10.1007/JHEP06(2017)067",
    journal = "JHEP",
    volume = "06",
    pages = "067",
    year = "2017"
}

@article{Katmadas:2018ksp,
    author = "Katmadas, S. and Ruggeri, D. and Trigiante, M. and Van Riet, T.",
    title = "{The holographic dual to supergravity instantons in $\rm AdS_5\times S^5/\mathbb{Z}_k$}",
    eprint = "1812.05986",
    archivePrefix = "arXiv",
    primaryClass = "hep-th",
    doi = "10.1007/JHEP10(2019)205",
    journal = "JHEP",
    volume = "10",
    pages = "205",
    year = "2019"
}

@article{VanRiet:2020pcn,
    author = "Van Riet, Thomas",
    title = "{Instantons, Euclidean wormholes and AdS/CFT}",
    eprint = "2004.08956",
    archivePrefix = "arXiv",
    primaryClass = "hep-th",
    doi = "10.22323/1.376.0121",
    journal = "PoS",
    volume = "CORFU2019",
    pages = "121",
    year = "2020"
}

@article{Hertog:2018kbz,
    author = "Hertog, Thomas and Truijen, Brecht and Van Riet, Thomas",
    title = "{Euclidean axion wormholes have multiple negative modes}",
    eprint = "1811.12690",
    archivePrefix = "arXiv",
    primaryClass = "hep-th",
    doi = "10.1103/PhysRevLett.123.081302",
    journal = "Phys. Rev. Lett.",
    volume = "123",
    number = "8",
    pages = "081302",
    year = "2019"
}

@article{Brown:2015iha,
    author = "Brown, Jon and Cottrell, William and Shiu, Gary and Soler, Pablo",
    title = "{Fencing in the Swampland: Quantum Gravity Constraints on Large Field Inflation}",
    eprint = "1503.04783",
    archivePrefix = "arXiv",
    primaryClass = "hep-th",
    reportNumber = "MAD-TH-15-04",
    doi = "10.1007/JHEP10(2015)023",
    journal = "JHEP",
    volume = "10",
    pages = "023",
    year = "2015"
}

@article{Giddings:1987cg,
    author = "Giddings, Steven B. and Strominger, Andrew",
    title = "{Axion Induced Topology Change in Quantum Gravity and String Theory}",
    reportNumber = "HUTP-87-A067",
    doi = "10.1016/0550-3213(88)90446-4",
    journal = "Nucl. Phys. B",
    volume = "306",
    pages = "890--907",
    year = "1988"
}

@article{Loges:2022nuw,
    author = "Loges, Gregory J. and Shiu, Gary and Sudhir, Nidhi",
    title = "{Complex saddles and Euclidean wormholes in the Lorentzian path integral}",
    eprint = "2203.01956",
    archivePrefix = "arXiv",
    primaryClass = "hep-th",
    doi = "10.1007/JHEP08(2022)064",
    journal = "JHEP",
    volume = "08",
    pages = "064",
    year = "2022"
}

@article{Aguilar-Gutierrez:2023ril,
    author = "Aguilar-Gutierrez, Sergio E. and Hertog, Thomas and Tielemans, Rob and van der Schaar, Jan Pieter and Van Riet, Thomas",
    title = "{Axion-de Sitter wormholes}",
    eprint = "2306.13951",
    archivePrefix = "arXiv",
    primaryClass = "hep-th",
    doi = "10.1007/JHEP11(2023)225",
    journal = "JHEP",
    volume = "11",
    pages = "225",
    year = "2023"
}

@article{Coleman:1988cy,
    author = "Coleman, Sidney R.",
    title = "{Black holes as red herrings: Topological fluctuations and the loss of quantum coherence}",
    reportNumber = "HUTP-88/A008",
    doi = "10.1016/0550-3213(88)90110-1",
    journal = "Nucl. Phys. B",
    volume = "307",
    pages = "867--882",
    year = "1988"
}

@article{Maldacena:2018lmt,
    author = "Maldacena, Juan and Qi, Xiao-Liang",
    title = "{Eternal traversable wormhole}",
    eprint = "1804.00491",
    archivePrefix = "arXiv",
    primaryClass = "hep-th",
    month = "4",
    year = "2018"
}

@article{Witten:1981gj,
    author = "Witten, Edward",
    title = "{Instability of the Kaluza-Klein Vacuum}",
    reportNumber = "PRINT-81-0441 (PRINCETON)",
    doi = "10.1016/0550-3213(82)90007-4",
    journal = "Nucl. Phys. B",
    volume = "195",
    pages = "481--492",
    year = "1982"
}

@article{OtherPaper,
    author = "Di Ubaldo, Gabriele and Iliesiu, Luca V. and Lin, Henry W. and Yan, Cynthia",
    title = "{Positivity of the gravitational path integral implies the axionic weak gravity conjecture}",
    eprint = "2605.05305",
    archivePrefix = "arXiv",
    primaryClass = "hep-th",
    reportNumber = "RIKEN-iTHEMS-Report-26",
    month = "5",
    year = "2026"
}

@article{Vafa:1994tf,
    author = "Vafa, Cumrun and Witten, Edward",
    title = "{A Strong coupling test of S duality}",
    eprint = "hep-th/9408074",
    archivePrefix = "arXiv",
    reportNumber = "HUTP-94-A017, IASSNS-HEP-94-54",
    doi = "10.1016/0550-3213(94)90097-3",
    journal = "Nucl. Phys. B",
    volume = "431",
    pages = "3--77",
    year = "1994"
}

@article{Bergshoeff:2004fq,
    author = "Bergshoeff, E. and Collinucci, Andres and Gran, U. and Roest, D. and Vandoren, S.",
    title = "{Non-extremal D-instantons}",
    eprint = "hep-th/0406038",
    archivePrefix = "arXiv",
    reportNumber = "KCL-MTH-04-07, SPIN-04-08, ITP-04-14, UG-04-02",
    doi = "10.1088/1126-6708/2004/10/031",
    journal = "JHEP",
    volume = "10",
    pages = "031",
    year = "2004"
}

@inproceedings{Stelle:1996tz,
    author = "Stelle, K. S.",
    title = "{Lectures on supergravity p-branes}",
    booktitle = "{ICTP Summer School in High-energy Physics and Cosmology}",
    eprint = "hep-th/9701088",
    archivePrefix = "arXiv",
    reportNumber = "IMPERIAL-TP-96-97-15",
    pages = "287--339",
    month = "6",
    year = "1996"
}

@article{Bergshoeff:2004pg,
    author = "Bergshoeff, E. and Collinucci, Andres and Gran, U. and Roest, D. and Vandoren, S.",
    editor = "Kiritsis, E.",
    title = "{Non-extremal instantons and wormholes in string theory}",
    eprint = "hep-th/0412183",
    archivePrefix = "arXiv",
    reportNumber = "KCL-MTH-04-16, ITP-UU-04-51, SPIN-04-33, UG-04-04",
    doi = "10.1002/prop.200410227",
    journal = "Fortsch. Phys.",
    volume = "53",
    pages = "990--996",
    year = "2005"
}

@article{Maldacena:2004rf,
    author = "Maldacena, Juan Martin and Maoz, Liat",
    title = "{Wormholes in AdS}",
    eprint = "hep-th/0401024",
    archivePrefix = "arXiv",
    reportNumber = "ITFA-2003-57",
    doi = "10.1088/1126-6708/2004/02/053",
    journal = "JHEP",
    volume = "02",
    pages = "053",
    year = "2004"
}

@article{Kontsevich:2021dmb,
    author = "Kontsevich, Maxim and Segal, Graeme",
    title = "{Wick Rotation and the Positivity of Energy in Quantum Field Theory}",
    eprint = "2105.10161",
    archivePrefix = "arXiv",
    primaryClass = "hep-th",
    doi = "10.1093/qmath/haab027",
    journal = "Quart. J. Math. Oxford Ser.",
    volume = "72",
    number = "1-2",
    pages = "673--699",
    year = "2021"
}

@article{Witten:2021nzp,
    author = "Witten, Edward",
    title = "{A Note On Complex Spacetime Metrics}",
    eprint = "2111.06514",
    archivePrefix = "arXiv",
    primaryClass = "hep-th",
    month = "11",
    year = "2021"
}

@article{Harlow:2022ich,
    author = "Harlow, Daniel and Heidenreich, Ben and Reece, Matthew and Rudelius, Tom",
    title = "{Weak gravity conjecture}",
    eprint = "2201.08380",
    archivePrefix = "arXiv",
    primaryClass = "hep-th",
    reportNumber = "ACFI-T22-01",
    doi = "10.1103/RevModPhys.95.035003",
    journal = "Rev. Mod. Phys.",
    volume = "95",
    number = "3",
    pages = "035003",
    year = "2023"
}

@article{Cheung:2014vva,
    author = "Cheung, Clifford and Remmen, Grant N.",
    title = "{Naturalness and the Weak Gravity Conjecture}",
    eprint = "1402.2287",
    archivePrefix = "arXiv",
    primaryClass = "hep-ph",
    reportNumber = "CALT-68-2879",
    doi = "10.1103/PhysRevLett.113.051601",
    journal = "Phys. Rev. Lett.",
    volume = "113",
    pages = "051601",
    year = "2014"
}

@article{Rudelius:2015xta,
    author = "Rudelius, Tom",
    title = "{Constraints on Axion Inflation from the Weak Gravity Conjecture}",
    eprint = "1503.00795",
    archivePrefix = "arXiv",
    primaryClass = "hep-th",
    doi = "10.1088/1475-7516/2015/9/020",
    journal = "JCAP",
    volume = "09",
    pages = "020",
    year = "2015"
}

@article{Montero:2015ofa,
    author = "Montero, Miguel and Uranga, Angel M. and Valenzuela, Irene",
    title = "{Transplanckian axions!?}",
    eprint = "1503.03886",
    archivePrefix = "arXiv",
    primaryClass = "hep-th",
    reportNumber = "IFT-UAM-CSIC-15-028, FTUAM-15-8",
    doi = "10.1007/JHEP08(2015)032",
    journal = "JHEP",
    volume = "08",
    pages = "032",
    year = "2015"
}

@article{Arkani-Hamed:2006emk,
    author = "Arkani-Hamed, Nima and Motl, Lubos and Nicolis, Alberto and Vafa, Cumrun",
    title = "{The String landscape, black holes and gravity as the weakest force}",
    eprint = "hep-th/0601001",
    archivePrefix = "arXiv",
    reportNumber = "HUTP-05-A0057",
    doi = "10.1088/1126-6708/2007/06/060",
    journal = "JHEP",
    volume = "06",
    pages = "060",
    year = "2007"
}

@article{Penington:2019kki,
    author = "Penington, Geoff and Shenker, Stephen H. and Stanford, Douglas and Yang, Zhenbin",
    title = "{Replica wormholes and the black hole interior}",
    eprint = "1911.11977",
    archivePrefix = "arXiv",
    primaryClass = "hep-th",
    doi = "10.1007/JHEP03(2022)205",
    journal = "JHEP",
    volume = "03",
    pages = "205",
    year = "2022"
}

@article{Saad:2018bqo,
    author = "Saad, Phil and Shenker, Stephen H. and Stanford, Douglas",
    title = "{A semiclassical ramp in SYK and in gravity}",
    eprint = "1806.06840",
    archivePrefix = "arXiv",
    primaryClass = "hep-th",
    month = "6",
    year = "2018"
}

@article{Almheiri:2019qdq,
    author = "Almheiri, Ahmed and Hartman, Thomas and Maldacena, Juan and Shaghoulian, Edgar and Tajdini, Amirhossein",
    title = "{Replica Wormholes and the Entropy of Hawking Radiation}",
    eprint = "1911.12333",
    archivePrefix = "arXiv",
    primaryClass = "hep-th",
    doi = "10.1007/JHEP05(2020)013",
    journal = "JHEP",
    volume = "05",
    pages = "013",
    year = "2020"
}

@article{Obers:1999um,
    author = "Obers, N. A. and Pioline, B.",
    title = "{Eisenstein series and string thresholds}",
    eprint = "hep-th/9903113",
    archivePrefix = "arXiv",
    reportNumber = "NORDITA-1999-18-HE, NBI-HE-99-06, CPHT-S710-0299",
    doi = "10.1007/s002200050022",
    journal = "Commun. Math. Phys.",
    volume = "209",
    pages = "275--324",
    year = "2000"
}

@article{Green:1981ya,
    author = "Green, Michael B. and Schwarz, John H.",
    title = "{Supersymmetrical Dual String Theory. 3. Loops and Renormalization}",
    reportNumber = "CALT-68-873",
    doi = "10.1016/0550-3213(82)90334-0",
    journal = "Nucl. Phys. B",
    volume = "198",
    pages = "441--460",
    year = "1982"
}

@article{Gross:1986iv,
    author = "Gross, David J. and Witten, Edward",
    title = "{Superstring Modifications of Einstein's Equations}",
    reportNumber = "Print-86-0250 (PRINCETON)",
    doi = "10.1016/0550-3213(86)90429-3",
    journal = "Nucl. Phys. B",
    volume = "277",
    pages = "1",
    year = "1986"
}

@article{Green:1997tv,
    author = "Green, Michael B. and Gutperle, Michael",
    title = "{Effects of D instantons}",
    eprint = "hep-th/9701093",
    archivePrefix = "arXiv",
    reportNumber = "DAMTP-96-104",
    doi = "10.1016/S0550-3213(97)00269-1",
    journal = "Nucl. Phys. B",
    volume = "498",
    pages = "195--227",
    year = "1997"
}

@article{Hertog:2024nys,
    author = "Hertog, T. and Maenaut, S. and Missoni, B. and Tielemans, R. and Van Riet, T.",
    title = "{Stability of axion-saxion wormholes}",
    eprint = "2405.02072",
    archivePrefix = "arXiv",
    primaryClass = "hep-th",
    doi = "10.1007/JHEP11(2024)151",
    journal = "JHEP",
    volume = "11",
    pages = "151",
    year = "2024"
}

@article{Andriolo:2020lul,
    author = "Andriolo, Stefano and Huang, Tzu-Chen and Noumi, Toshifumi and Ooguri, Hirosi and Shiu, Gary",
    title = "{Duality and axionic weak gravity}",
    eprint = "2004.13721",
    archivePrefix = "arXiv",
    primaryClass = "hep-th",
    reportNumber = "CALT-TH 2020-007, IPMU20-0035, KOBE-COSMO-20-04, MAD-TH-20-02",
    doi = "10.1103/PhysRevD.102.046008",
    journal = "Phys. Rev. D",
    volume = "102",
    number = "4",
    pages = "046008",
    year = "2020"
}

@article{Kats:2006xp,
    author = "Kats, Yevgeny and Motl, Lubos and Padi, Megha",
    title = "{Higher-order corrections to mass-charge relation of extremal black holes}",
    eprint = "hep-th/0606100",
    archivePrefix = "arXiv",
    reportNumber = "HUTP-06-A0023",
    doi = "10.1088/1126-6708/2007/12/068",
    journal = "JHEP",
    volume = "12",
    pages = "068",
    year = "2007"
}

@article{Loveridge:2025dls,
    author = "Loveridge, Andrew and Sun, Hao-Yu",
    title = "{AdS$_{3}$ axion wormholes as stable contributions to the Euclidean gravitational path integral}",
    eprint = "2504.10868",
    archivePrefix = "arXiv",
    primaryClass = "hep-th",
    doi = "10.1007/JHEP04(2026)138",
    journal = "JHEP",
    volume = "04",
    pages = "138",
    year = "2026"
}
\bibliographystyle{JHEP.bst}
\end{document}